\documentclass[a4paper,superscriptaddress,11pt,twoside]{revtex4}

\usepackage{enumerate}
\usepackage{graphics}
\usepackage{multirow}
\usepackage{color}
\usepackage{lscape,graphicx,epic,eepic,epsfig,amsmath,latexsym,amssymb,revsymb,color,fancyhdr}
\usepackage{times}
\usepackage{textcomp}

\newtheorem{theorem}{Theorem}
\usepackage{lscape}
\newtheorem{lemma}[theorem]{Lemma}

\newtheorem{proposition}[theorem]{Proposition}

\newtheorem{corollary}[theorem]{Corollary}

\newcommand\ket[1]{\ensuremath{|#1\rangle}}
\newcommand\bra[1]{\ensuremath{\langle#1|}}
\newcommand\proj[1]{\ensuremath{|#1\rangle \langle#1|}}

\def\Tr{\operatorname{tr}}
\def\>{\rangle}
\def\<{\langle}

\def\be{\begin{equation}}
\def\ee{\end{equation}}
\def\ben{\begin{eqnarray}}
\def\een{\end{eqnarray}}
\def\ot{\otimes}
\def\LOCC{\text{LOCC}}
\def\ALL{\text{ALL}}

\def\squareforqed{\hbox{\rlap{$\sqcap$}$\sqcup$}}
\def\qed{\ifmmode\squareforqed\else{\unskip\nobreak\hfil
\penalty50\hskip1em\null\nobreak\hfil\squareforqed
\parfillskip=0pt\finalhyphendemerits=0\endgraf}\fi}
\def\endenv{\ifmmode\;\else{\unskip\nobreak\hfil
\penalty50\hskip1em\null\nobreak\hfil\;
\parfillskip=0pt\finalhyphendemerits=0\endgraf}\fi}

\newlength{\blank}
\newenvironment{proof}[1][{\hspace{-\blank}}]{{\noindent\textbf{Proof   }}}{\hfill\end{proof}\vskip 0.5\baselineskip}

\mathchardef\ordinarycolon\mathcode`\:
\mathcode`\:=\string"8000
\def\vcentcolon{\mathrel{\mathop\ordinarycolon}}
\begingroup \catcode`\:=\active
  \lowercase{\endgroup
  \let :\vcentcolon
  }

\newcommand{\nc}{\newcommand}
\nc{\rnc}{\renewcommand}
\nc{\beq}{\begin{equation}}
\nc{\eeq}{{\end{equation}}}
\nc{\beqa}{\begin{eqnarray}}
\nc{\eeqa}{\end{eqnarray}}
\nc{\lbar}[1]{\overline{#1}}
\nc{\ketbra}[2]{|#1\rangle\!\langle#2|}
\nc{\braket}[2]{\langle#1|#2\rangle}
\nc{\avg}[1]{\langle#1\rangle}
\nc{\Rank}{\operatorname{rank}\,}
\nc{\smfrac}[2]{\mbox{$\frac{#1}{#2}$}}
\nc{\tr}{\mathrm{Tr}}
\nc{\ox}{\otimes}
\nc{\dg}{\dagger}
\nc{\dn}{\downarrow}
\nc{\cA}{{\cal A}}
\nc{\cB}{{\cal B}}
\nc{\cC}{{\cal C}}
\nc{\cD}{{\cal D}}
\nc{\cE}{{\cal E}}
\nc{\cF}{{\cal F}}
\nc{\cG}{{\cal G}}
\nc{\cH}{{\cal H}}
\nc{\cI}{{\cal I}}
\nc{\cJ}{{\cal J}}
\nc{\cK}{{\cal K}}
\nc{\cL}{{\cal L}}
\nc{\cM}{{\cal M}}
\nc{\cN}{{\cal N}}
\nc{\cO}{{\cal O}}
\nc{\cP}{{\cal P}}
\nc{\cR}{{\cal R}}
\nc{\cS}{{\cal S}}
\nc{\cT}{{\cal T}}
\nc{\cU}{{\cal U}}
\nc{\cX}{{\cal X}}
\nc{\cZ}{{\cal Z}}
\nc{\csupp}{{\operatorname{csupp}}}
\nc{\qsupp}{{\operatorname{qsupp}}}
\nc{\var}{\operatorname{var}}
\nc{\rar}{\rightarrow}
\nc{\lrar}{\longrightarrow}
\nc{\polylog}{\operatorname{polylog}}
\nc{\1}{{\openone}}

\def\a{\alpha}
\def\b{\beta}
\def\g{\gamma}
\def\d{\delta}
\def\e{\epsilon}

\def\m{\mu}
\def\n{\nu}

\def\r{\rho}
\def\s{\sigma}

\def\G{\Gamma}

\def\P{\Pi}

\def\Ps{\Psi}

\nc{\RR}{{{\mathbb R}}}
\nc{\CC}{{{\mathbb C}}}
\nc{\FF}{{{\mathbb F}}}
\nc{\NN}{{{\mathbb N}}}
\nc{\ZZ}{{{\mathbb Z}}}
\nc{\PP}{{{\mathbb P}}}
\nc{\QQ}{{{\mathbb Q}}}
\nc{\UU}{{{\mathbb U}}}
\nc{\EE}{{{\mathbb E}}}
\nc{\id}{{\operatorname{id}}}

\nc{\bea}{\begin{eqnarray}}
\nc{\eea}{\end{eqnarray}}
\nc{\Hom}[2]{\mbox{Hom}(\CC^{#1},\CC^{#2})}
\nc{\rU}{\mbox{U}}

\nc{\ob}[1]{#1}

\nc{\LO}{\text{LO}}
\nc{\cLOCC}{{\overline{\text{LOCC}}}}
\nc{\SEP}{\text{SEP}}
\nc{\PPT}{\text{PPT}}
\nc{\ccq}{\text{ccq}}
\nc{\sep}{\text{sep}}
\nc{\twist}{\text{twist}}

\nc{\ABC}{{A\leftrightarrow C\leftrightarrow B}}
\nc{\ABCr}{{A\leftarrow C\rightarrow B}}
\nc{\ABCrr}{{A\leftarrow C\leftrightarrow B}}
\nc{\CAB}{{C\leftrightarrow AB}}
\nc{\CABr}{{C\rightarrow AB}}

\nc{\te}{\otimes}

\newcommand*{\pro}[1]{\ket{#1}\!\bra{#1}}

\begin{document}

\title{Limitations on Quantum Key Repeaters}

\author{Stefan B{\"a}uml}
\email{stefan.baeuml@bristol.ac.uk}
\affiliation{Department of Mathematics, University of Bristol, Bristol BS8 1TW, UK}
\affiliation{F\'{\i}sica Te\`{o}rica: Informaci\'{o} i Fenomens  Qu\`{a}ntics, Universitat Aut\`{o}noma de Barcelona, ES-08193 Bellaterra (Barcelona), Spain}
              
\author{Matthias Christandl}
\email{christandl@math.ku.dk}
\affiliation{Department of Mathematical Sciences, University of Copenhagen, Universitetsparken 5, 2100 Copenhagen, Denmark}

\author{Karol Horodecki}
\email{khorodec@inf.ug.edu.pl}
\affiliation{Institute of Informatics, University of Gda\'nsk, 80-952 Gda\'nsk, Poland}
\affiliation{National Quantum Information Centre of Gda\'nsk, 81-824 Sopot, Poland}

\author{Andreas Winter}
\email{andreas.winter@uab.cat}
\affiliation{ICREA - Instituci\'{o} Catalana de Recerca i Estudis Avan\c{c}ats, ES-08010 Barcelona, Spain}
\affiliation{F\'{\i}sica Te\`{o}rica: Informaci\'{o} i Fenomens Qu\`{a}ntics, Universitat Aut\`{o}noma de Barcelona, ES-08193 Bellaterra (Barcelona), Spain}
\affiliation{Department of Mathematics, University of Bristol, Bristol BS8 1TW, UK}



\begin{abstract}
A major application of quantum communication is the distribution of entangled 
particles for use in quantum key distribution (QKD). 
Due to noise in the communication line, QKD is in practice limited to a distance of a few hundred kilometres, and can only be extended to longer distances by use of a quantum 
repeater, a device which performs entanglement 
distillation and quantum teleportation. 
The existence of noisy entangled states that are undistillable but nevertheless 
useful for QKD raises the question of the feasibility of a quantum key repeater, which would work beyond the limits of entanglement 
distillation, hence possibly tolerating higher noise levels than existing protocols. Here we exhibit fundamental limits on such a device in the form of bounds on the rate at which it may extract secure key. As a consequence, we give examples of states suitable for QKD but unsuitable for the most general quantum key repeater protocol. 
\end{abstract}

\maketitle

When a signal is passed from a sender to a receiver, it inevitably degrades due to the noise present in any realistic communication channel (for example~a cable or free space). The degradation of the signal is typically exponential in the length of the communication line. When the signal is classical, degradation can be counteracted by use of an amplifier that measures the degraded signal and, depending on a threshold, replaces it by a stronger signal. When the signal is quantum mechanical (for example encoded in non-orthogonal polarisations of a single photon), such an amplifier cannot work any more, since the measurement inevitably disturbs the signal \cite{fuchs-disturbance}, and, more generally, since quantum mechanical signals cannot be cloned \cite{no-cloning}. Sending a quantum signal, however, is the basis of quantum key distribution (QKD), a method to distribute a cryptographic key which can later be used for perfectly secure communication between sender and receiver \cite{BB84}. The degradation of sent quantum signals therefore seems to place a fundamental limit on the distance at which secure communication is possible thereby severely limiting its applicability in the internet \cite{stucki2009high, gisin-review,scarani2009security}.

A way around this limitation is the use of entanglement-based quantum key distribution schemes \cite{EK91,BBM92} in conjunction with a so-called quantum repeater \cite{repeatersPRL,SSRG11}.
This amounts to distributing $n$ Einstein-Podolsky-Rosen (EPR) pairs between Alice and Charlie (an untrusted telecom provider) and between Bob and Charlie. Imperfections due to noise in the transmission are compensated by distillation, yielding $\approx E_\text{D} \times n$ perfect EPR pairs. Here $E_\text{D}$ denotes the distillable entanglement of the imperfect EPR pair, that is the optimal rate at which perfect EPR pairs can be distilled from imperfect ones.
The EPR pairs between Charlie and Bob are then used to teleport the state of Charlie's other particles to Bob. This process, known as entanglement swapping, results in EPR pairs between Alice and Bob \cite{ZZHE93} (see Fig. 1). When Alice and Bob make appropriate measurements on these EPR pairs, they obtain a sequence of secret key bits, that is, an identical but random sequence of bits that is uncorrelated with the rest of the universe (including Charlie's systems), enabling secure communication.
The described scheme with one intermediate station effectively doubles the distance over which QKD can be carried out. This abstract view of the quantum repeater will be sufficient for our purpose. The full proposal of a quantum repeater in fact allows to efficiently extend the distance arbitrarily even if the local operations are subject to a limited amount of noise \cite{repeatersPRL}. The implementation of quantum repeaters is therefore 
one of the focal points of experimental quantum information science \cite{SSRG11}. 

\begin{figure}	
	\centering
	\includegraphics[trim=4cm 7cm 2cm 0cm,clip=true,width=0.6\textwidth]{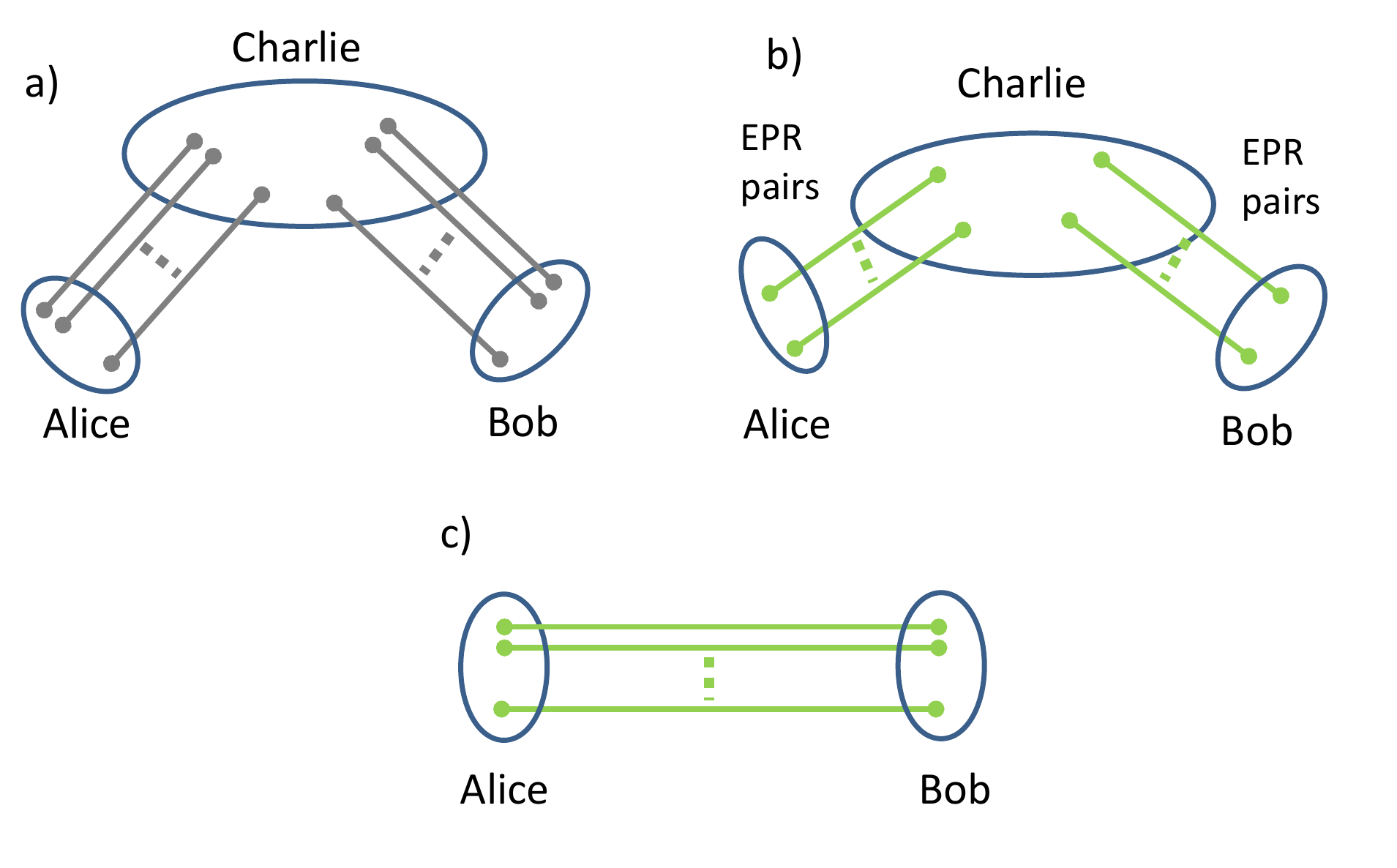}
	\caption{Quantum repeater: a) Alice and Charlie -- and similarly Charlie and Bob -- distil EPR pairs from noisy states (grey). b) Charlie uses the EPR pairs (green) he shares with Bob to teleport his part of the states he shares with Alice to Bob. c) Alice and Bob share EPR pairs. }
 
	\label{fig:q-rep}
\end{figure}


Due to the tight connection between the distillation of EPR pairs and QKD \cite{quantum-privacy-amplification,shor-preskill}, it came as a surprise that there are bound entangled states (that is entangled states with vanishing distillable entanglement) from which secret key can be obtained \cite{pptkey}. With the help of a quantum repeater as described above, however, the secret key contained in such states cannot be extended to larger distances, as the states do not allow for the distillation of EPR pairs. This raises the question of whether there may be other ways to extend the secret key to arbitrary distances than by entanglement distillation and swapping, other quantum key repeaters. 

In this work, we introduce and formally define the concept of a quantum key repeater. We then study the associated quantum key repeater rate. It is always at least as large as the rate that can be obtained in a quantum repeater protocol and we raise the question whether it could be larger (and in particular non-zero for bound entangled states). Our main results consist of upper bounds on this quantity which we use to show that there are quantum states with extreme behaviour: state with a large key rate but with a  negligible quantum key repeater rate. We thus demonstrate fundamental limitations on quantum key repeaters. 

\section*{Results}
{\it The Quantum Key Repeater Rate}

We analyse the quantum key repeater rate $K_{\text{A}\leftrightarrow \text{C} \leftrightarrow \text{B}}$ at which a protocol --- only using local operations and classical communication (LOCC) --- is able to extract private bits between Alice and Bob from entangled states which each of them shares with Charlie (see Fig. 2). See Supplementary Note 1 for a formal definition of the key repeater rate. By a private bit we mean an entangled state containing a unit of privacy paralleling the EPR pair as a unit of entanglement \cite{pptkey,keyhuge}. Mathematically, private bits are entangled states of the form
\begin{eqnarray}
\gamma_{{\text{AA'}\text{BB'}}}=\frac{1}{2}\left[ \begin{array}{cccc}
\sqrt{X X^{\dagger}} & 0 & 0 & X \\
0 & 0& 0 & 0 \\
0 & 0 & 0 & 0\\
X^{\dagger} & 0 & 0 & \sqrt{X^{\dagger} X}\\
\end{array}
\right],
\end{eqnarray}
where $A$ and $B$ are qubits that contain the key bits, corresponding to the rows and columns in the matrix. The AB subsystem is called the key part.  A' and B' are each a $d$-dimensional systems, forming the so-called shield part. $X$ is a $d^2$-by-$d^2$ matrix with $\|X\|_1=1$ (see also Fig. 3). $\gamma_{\text{AA'}\text{BB'}}$ can also be presented in the form $U \proj{\Psi}_\text{AB} \otimes \sigma_{\text{A'}\text{B'}} U^{\dagger}$, where $\sigma_{\text{A'}\text{B'}}$ is some state, $\ket{\Psi}=\frac{1}{\sqrt{2}} \ket{00+11}$ and $U=\proj{00}_\text{AB}\otimes U_0+\proj{11}_\text{AB}\otimes U_1$ is a controlled unitary acting on $\sigma_{\text{A'}\text{B'}}$. This operation is called twisting. It is now easy to see that the bit that Alice and Bob obtain when they measure $A$ and $B$ in the computation basis is a key bit, that is, it is random and secure, that is product with a purification of $\gamma$ held by the eavesdropper. The relation between $X$ and $\sigma$ is given by $X=U_0 \sigma_{\text{A'}\text{B'}} U_1^\dagger$. 


\begin{figure}
	\centering
	\includegraphics[trim=2cm 10cm 0cm 0cm,clip=true,width=0.9\textwidth]{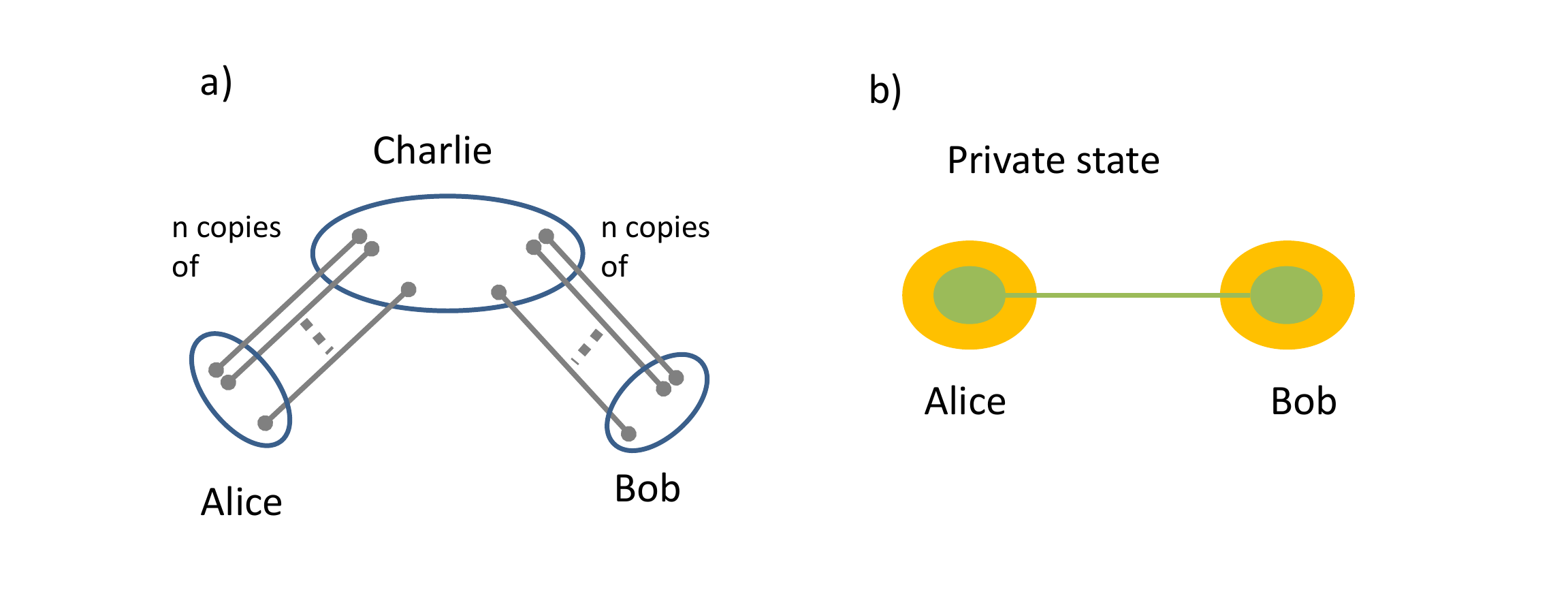}
	\caption{Quantum key repeater: a) Multiple copies of noisy states $\rho$ and $\tilde\rho$, shared by Alice and Charlie and by Charlie and Bob, respectively, are transformed by means of LOCC into b) a private state $\gamma$ (green-yellow) between Alice and Bob.}
	\label{fig:key-rep}
\end{figure}

Note that just as the definition of the distillable key \cite{DevetakWinter-hash,pptkey}, the definition of the quantum key repeater rate is information-theoretic in nature. 
The role of Charlie here merits special attention. While he participates in 
the LOCC protocol like Alice and Bob do, he is not a ``trusted party''; 
indeed, at the end of the protocol, Alice and Bob wish to obtain private bits, whose privacy is not compromised even if at that point Charlie passes all his remaining information to the eavesdropper.
We also note that well-known techniques from quantum information theory \cite{post-selection,privacyamplification} allow to conclude that the obtained rate of private bits can be made unconditionally secure \cite{BenOrUniversal, unruh,HHHLO:unco-pbit}. In the following we will describe our main results which demonstrate that the 
performance of quantum key repeaters beyond the use of entanglement distillation is severely limited.
\begin{figure}
	\centering
	\includegraphics[trim=3cm 5cm 4cm 2.5cm,clip=true,width=0.6\textwidth]{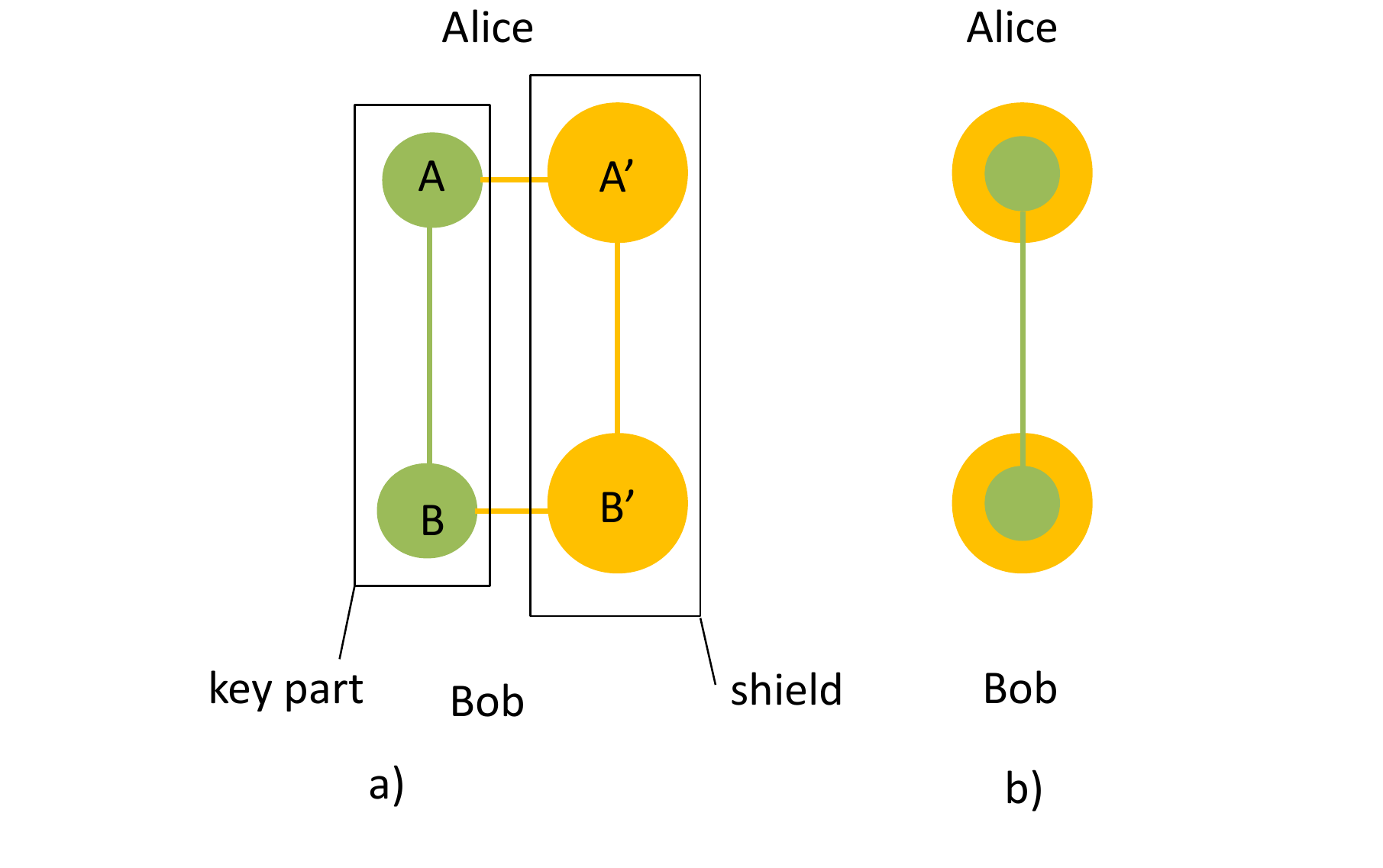}
	\caption{The private state $\gamma_{\text{AA'}\text{BB'}}$. a) Bipartite state with four subsystems A,A',B and B'. The subsystems AB form the "key part" (green) which, due to the "shield part" A'B' (yellow), is secure against an eavesdropper. b) Icon of a private bit.}
	\label{fig:pbit}
\end{figure}

{\it Some private states cannot be swapped}

Our first result takes as its starting point the observation that there are private bits that are almost indistinguishable from separable states by LOCC \cite{karol-PhD}. To see this, consider the state 
\begin{eqnarray}
\tilde\gamma_{\text{AA'}\text{BB'}}=\frac{1}{2}\left[ \begin{array}{cccc}
\sqrt{X X^{\dagger}} & 0 & 0 & 0 \\
0 & 0& 0 & 0 \\
0 & 0 & 0 & 0\\
0 & 0 & 0 & \sqrt{X^{\dagger} X}\\
\end{array}
\right],
\end{eqnarray}
which is obtained from $\gamma$, when Alice and Bob measure the key part of their state in the computational basis. An example is given by the choice $X=\frac{1}{d\sqrt{d}}\sum_{ij} u_{ij}\ket{i}\bra{j}\ot \ket{j}\bra{i}$, where the $u_{ij}$ are the entries in the quantum Fourier transform in dimension $d$. For this choice of $X$, $\tilde\gamma$ is separable. The distinguishability under LOCC operations is measured in the norm $\|\gamma-\tilde\gamma\|_{\text{LOCC}}$, which is bounded by the distinguishability under global maps preserving the positivity under the partial transpose $\|\gamma-\tilde\gamma\|_{\text{PPT}}$ \cite{EW02}. This can further be bounded by $ \|\gamma^\Gamma-\tilde\gamma^\Gamma\|_{1}$, which is easily calculated as $\|X^\Gamma\|_1=\frac{1}{\sqrt{d}+1}$. $\Gamma$ indicates the partial transpose, that is, the transpose of one of the systems \cite{smallkey}. 

Suppose now that a quantum repeater protocol applied to two copies of the latter state, shared by Alice and Charlie and Bob and Charlie respectively, successfully outputs a private bit between Alice and Bob. This could be regarded as the privacy analogue to entanglement swapping.
Then, if Alice and Bob joined their labs, they could distinguish this resulting state from a separable state,
as separable states are well distinguishable from private states by a global measurement \cite{pptkey}.
This implies an LOCC procedure for Alice \&{} Bob (jointly) and Charlie to distinguish the initial private bits $\gamma\ot\gamma$ from separable states: first run the quantum key repeater protocol and then perform the measurement. This, however, is in contradiction to the property that the private state $\gamma$ (and hence $\gamma \ot \gamma$) is almost indistinguishable from separable states under LOCC. In conclusion this shows that such private bits cannot be successfully extended to a private bit between Alice and Bob by any LOCC protocol acting on single copies (see Supplementary Note 2).

{\it Bounding the Quantum Key Repeater Rate}

Although intuitive, the above argument only bounds the repeated key obtained from a 
{\it single} copy of input states.
%
The language of entanglement measures allows us to formulate this argument asymptotically as a rigorous distinguishability bound on the rate $K_{\text{A}\leftrightarrow \text{C} \leftrightarrow \text{B}}$ for general states $\rho$ and $\tilde\rho$:
\be
  K_{\text{A} \leftrightarrow \text{C} \leftrightarrow \text{B}}(\rho_{\text{AC}_\text{A}}\otimes \tilde\rho_{\text{C}_\text{B}\text{B}})
    \leq D_{\text{C}\leftrightarrow\text{AB}}^\infty(\rho_{\text{AC}_\text{A}}\otimes \tilde\rho_{\text{C}_\text{B}\text{B}}),
\ee
where the right hand side is the regularised LOCC-restricted relative entropy distance to the closest separable state \cite{Piani2009-relent}: $D^\infty(\rho)=\lim_{n\mapsto \infty} \frac{1}{n}D(\rho^{\otimes n})$, where $D(\r)=\inf_{\sigma}\sup_{M}D(M(\rho)||M(\sigma))$ with the minimisation over separable states $\sigma$, the maximisation over LOCC implementable measurements and $D$ the relative entropy distance. The proof is given in Supplementary Note 3.

Arguably, it is difficult if not impossible to compute this expression. But noting that this bound is invariant under partial transposition of the $C$ system, we can easily upper bound the quantity for all known bound entangled states (these are the ones with positive partial transpose) in terms of the relative entropy of entanglement of the partially transposed state $\rho^\Gamma$: 
$E_\text{R}^\infty(\rho^\Gamma)+E_\text{R}^\infty(\tilde\rho^\Gamma)$. The relative entropy of entanglement is given by $E_\text{R}(\rho)=\min_\sigma D(\rho||\sigma) $ where the minimisation extends over separable states; the regularisation is analogous to the one above. If we restrict to forward communication from Charlie and $\rho_{\text{AC}_\text{A}}=\tilde\rho_{\text{C}_\text{B}\text{B}}$, the squashed entanglement measure provides a bound: $K_{\text{A} \leftarrow\text{C} \rightarrow \text{B}}(\rho_{\text{AC}_\text{A}}\otimes \tilde\rho_{\text{C}_\text{B}\text{B}})\leq 4E_\text{sq}(\rho^\Gamma)$. The squashed entanglement is given as (one half times) the minimal conditional mutual information when minimising over all extensions of the state (we condition on the extending system).
Using invariance under partial transposition directly on the hypothetical
quantum key repeater protocol, we  
obtain for PPT states $\rho$ and $\tilde\rho$: 
\be
  K_{\text{A} \leftrightarrow \text{C} \leftrightarrow \text{B}}(\rho_{\text{AC}_\text{A}}\otimes \tilde\rho_{\text{C}_\text{B}\text{B}})
     \leq K_\text{D}(\rho_{\text{AC}_\text{A}}^\Gamma) \leq \min\{E_\text{R}^\infty(\rho_{\text{AC}_\text{A}}^{\Gamma}), E_\text{sq}(\rho_{\text{AC}_\text{A}}^{\Gamma})\},
\ee 
where $K_\text{D}$ is the key rate, that is, the rate at which secret key can be extracted from $\rho$ by LOCC. The same holds for $\tilde{\rho}_{\text{C}_\text{B}\text{B}}^{\Gamma}$. The proof can be found in Supplementary Note 4.

We will now give an example of a state $\rho_{\text{AC}_\text{A}}=\tilde\rho_{\text{C}_\text{B}\text{B}}$ for which the key rate is large, but the bounds, hence the quantum key repeater rate, are arbitrarily small. 
Guided by our intuition, we would like to consider the private bit $\gamma$ from above whose partial transpose is close to a separable state. The state, however, is not PPT, as no private bit can be PPT \cite{pptkey}. Fortunately, it turns into a PPT state $\rho$ under mixing with a small amount of noise and we find $K_{\text{A} \leftrightarrow \text{C} \leftrightarrow \text{B}}(\rho \ot \rho )\approx 0 $ while $K_\text{D}(\rho) \approx 1$. 
This leads us to the main conclusion of our paper: there exist entangled quantum states that are useful for quantum key distribution at small distances but that are virtually useless for long-distance quantum key distribution (see Fig. 4).

\begin{figure}
	\centering
	\includegraphics[trim=0cm 0cm 5.5cm 0cm,clip=true,width=0.6\textwidth]{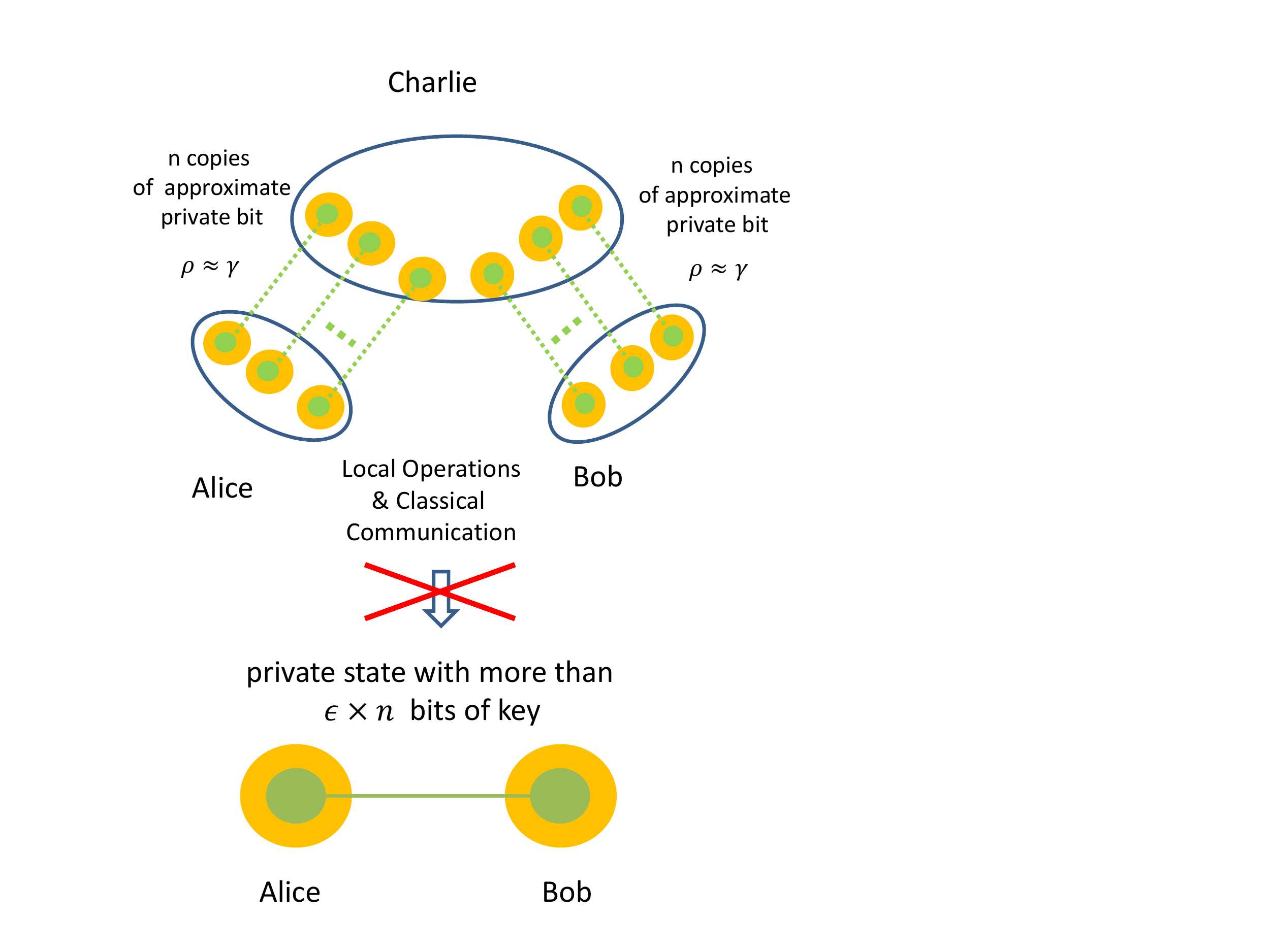}
	\caption{Limitation on quantum key repeaters: Despite Alice and Charlie as well as Charlie and Bob sharing almost n bits
of secure key, there is no LOCC protocol between Alice, Charlie and Bob, which results in a non-negligible amount of secure key between Alice and Bob.}
	\label{fig:nogo}
\end{figure}
{\it Bounding the Entanglement of the Output}

Finally, we present a different type of bound on the quantum key repeater rate based on the direct analysis of the entanglement of a concrete output state of a quantum repeater protocol:
\begin{equation} \label{eq:bound2}K_{\text{A} \leftarrow \text{C} \leftrightarrow \text{B}}(\rho_{\text{AC}_\text{A}} \ot \tilde\rho_{\text{C}_\text{B}\text{B}}) \leq \frac{1}{2}E_\text{C}(\rho_{\text{AC}_\text{A}}) +\frac{1}{2}E_\text{D}(\tilde\rho_{\text{C}_\text{B}\text{B}}),\end{equation}
where $E_\text{C}$ denotes the entanglement cost of the state, the rate of EPR states needed to create many copies of the state.
This bound, unlike the ones presented above, applies to all quantum states.
In particular, it applies to certain states invariant under partial transposition which escape the techniques presented before.
Note that in the case of PPT states, one may partially transpose the states appearing on the right hand side since $K_{\text{A} \leftarrow \text{C} \leftrightarrow \text{B}}$ is invariant under partial transposition. The proof of \eqref{eq:bound2} is obtained by upper bounding the squashed entanglement of the output state of the protocol using a manipulation of entropies resulting in the right hand side of \eqref{eq:bound2}. The squashed entanglement in turn upper bounds the distillable key of the output state (which upper bounds the left hand side) \cite{Christandl-Schuch-Winter}. For a detailed proof see Supplementary Note 5. There, we also exhibit a private bit with a significant drop in the repeater rate when compared to the key rate. We further investigate the tightness of the bound \eqref{eq:bound2} and, based on a random construction, show that the left hand side cannot be replaced by the entanglement cost of the output state. 
\section*{Discussion}

The preceding results pose limitations on the entanglement of the output state of a quantum key repeater protocol. As such, they support the PPT-squared conjecture: Assume that Alice and Charlie share a PPT state and that Bob and Charlie share a PPT state; then the state of Alice and Bob, conditioned on any measurement by Charlie, is always separable \cite{pptsquare,Dipl,master-hansen}.
Reaching even further, and consistent with our findings, we may speculate that perhaps the only ``transitive'' entanglement in quantum states, that is~entanglement that survives a quantum key repeater, is the distillable entanglement. One may also wonder whether apart from (\ref{eq:bound2}) there are other inequalities between entanglement measures of the in- and output states. In the context of algebro-geometric measures, this question has been raised and relations for the concurrence have been found \cite{gour-assistance, lee2011distribution}. Our work focuses on operational entanglement measures. 

States from which more key than entanglement can be extracted have recently been demonstrated experimentally in a quantum optical setup \cite{dobekPRL}. These are exactly the private states discussed in Supplementary Note 2 ($X$ is the SWAP operator) with shield dimension equal to two. As our results for these states only become effective for higher shield dimensions, we cannot conclude that the single copy key repeater drops when compared to the key contained in these states. This may be overcome by stronger theoretical bounds or experimental progress which increases the shield dimension; we expect both improvements to be achieved in the near future. 

With this paper we initiate the study of long-distance quantum communication and cryptography beyond the use of entanglement distillation by the introduction of the concept of a quantum key repeater. Even though the reported results provide limitations rather than new possibilities, we hope that this work will lead to a rethinking of the currently used protocols resulting in procedures for long-distance quantum communication that are both more efficient and that can operate in noisier environments. In the following we will give a simple example of such a rethinking: Assume that Alice and Charlie share a private bit $\gamma_{\text{AC}_\text{A}}$ which is almost PPT and thus requires a large shield system (see Supplementary Note 6). The quantum repeater based on quantum teleportation would thus require Bob and Charlie to share a large amount of EPR pairs in order to teleport Charlie's share of $\gamma_{\text{AC}_\text{A}}$ to Bob. Alice and Bob can then extract one bit of secret key by measuring the state. Inspired by the work of Smith and Yard \cite{SmithYard}, we show in Supplementary Note 6 that a single EPR pair and a particular state $\rho_{\text{C}_\text{B}\text{B}}$ which is so noisy that it contains no (one-way) distillable entanglement are sufficient in order to obtain a large quantum key repeater rate (using only one-way communication from Alice and Charlie to Bob). We thus showed that there are situations in which significant amounts of distillable entanglement may be replaced by (one-way) undistillable states.

\section*{References}


\section*{Achknowledgements}
Part of this work was done when the authors attended the
program ``Mathematical Challenges in Quantum Information'', Aug-Dec 2013, 
at the Isaac Newton Institute for Mathematical Sciences, Cambridge,
whose hospitality is gratefully acknowledged. MC was with ETH Zurich and visiting the Centre for Quantum Information and Foundations, DAMTP, University of Cambridge, during part of this work. We thank Gl\'{a}ucia Murta for pointing out an error in an earlier version of the manuscript. KH thanks Micha\l{} and Pawe\l{} Horodecki and Jonathan Oppenheim for helpful discussions.

MC was supported by a DFF Sapere Aude grant, an ERC Starting Grant, the CHIST-ERA project ``CQC'', an SNSF Professorship, the Swiss NCCR ``QSIT'', and the Swiss SBFI in relation to COST action MP1006. KH acknowledges support by the ERC Advanced Grant ``QOLAPS'' and the National Science Centre project Maestro DEC-2011/02/A/ST2/00305. AW was supported by the Spanish MINECO, projects FIS2013-40627-P and FIS2008-01236 with the support of FEDER funds, the Generalitat de Catalunya CIRIT project 2014 SGR 966, the EC STREP ``RAQUEL'', and the Philip Leverhulme Trust. AW and SB were supported by the ERC Advanced Grant ``IRQUAT''.
\pagebreak

\section*{Supplementary Note 1}\label{sec:intro}
\subsection*{Definitions}
Here we first formally recall the definition of a private state, of the secret key rate and of the distillable entanglement. We will then introduce the distillation of secure key with an intermediate station and formally introduce the corresponding information theoretic rate of secure key. A private state can be constructed from a maximally entangled state $\ket{\Psi^{2^m}}_{AB}=\sum_{i=0}^{2^m-1} \ket{ii}=\ket{\Psi}^{\ot m}$ by tensoring with some state $\s_{A'B'}$ and performing a so-called ``twisting`` operation. A twisting operation is a controlled unitary of the form $U^{\twist}=\sum_{ij}{\pro{ij}_{AB}\otimes U^{(ij)}_{A'B'}}$ that spreads the entanglement over the enlarged Hilbert space. Formally 
\begin{align}
	\g_{m} &=U^{\twist}\left(\pro{\Ps^{(2^m)}}_{AB}\otimes\s_{A'B'}\right){U^{\twist}}^\dagger \\
        &=\frac{1}{2^m}\sum_{ij=0}^{2^m-1}{\ket{ii}\bra{jj}_{AB}\otimes U^{(ii)}\s_{A'B'}{U^{(jj)}}^\dagger},
\end{align}
where we emphasize that $m$ is the number of key bits, in contrast to some of the literature, where the subscript denotes the dimension of the key system. It has been shown that even if Eve is in possession of the entire 
purification of $\g_m$, Alice and Bob will still be able to 
obtain $m$ bits of perfect key by measuring the $AB$ subsystem 
in the computational basis, while keeping the $A'B'$ part  away from 
Eve. As all the correlation the key has with the outside world is 
contained in $A'B'$, it is called the ``shield part``, whereas $AB$ 
is called the ``key part``. For $m=1$, $\g_1$ is also called a ``private bit`` 
or ``p-bit`` which can alternatively be represented in the form 
\begin{eqnarray}
\g_1^{AA'BB'}=\frac{1}{2}\left[ \begin{array}{cccc}
\sqrt{X X^{\dagger}} & 0 & 0 & X \\
0 & 0& 0 & 0 \\
0 & 0 & 0 & 0\\
X^{\dagger} & 0 & 0 & \sqrt{X^{\dagger} X}\\
\end{array}
\right],
\label{eq:example-gamma}
\end{eqnarray}
where $A$ and $B$ are qubits that contain the key bits, corresponding to the rows and columns in the matrix. $A'$ and $B'$ are each $d$-dimensional systems, called the shield. $X$ is a $d^2$-by-$d^2$ matrix with $\|X\|_1=1$.
 As the twisting operations can be non-local, not every private state can be obtained 
from a single rank $2^m$ maximally entangled state via LOCC. This shows that privacy is a 
truly different property of a quantum state than its distillable 
entanglement, motivating the introduction of a quantity known as 
``distillable key`` \cite{pptkey} 
\begin{equation}
  \label{KD}
  K_D(\rho) = \inf_{\epsilon > 0} \limsup_{n \rightarrow \infty} 
                \sup_{\Lambda_n \text{ LOCC}, \gamma_m} 
              \left\{ \frac{m}{n}: \Lambda_n(\rho^{\otimes n})\approx_\epsilon \gamma_m \right\},
\end{equation}
in analogy to the distillable entanglement
\begin{equation}
  \label{ED}
  E_D(\rho) = \inf_{\epsilon > 0} \limsup_{n \rightarrow \infty} 
                \sup_{\Lambda_n \text{ LOCC}} 
\left\{ \frac{m}{n}: \Lambda_n(\rho^{\otimes n}) \approx_\epsilon \pro{\Psi}^{\ot m} \right\}.
\end{equation}
With $\alpha\approx_\epsilon \beta$ we mean $\|\alpha-\beta \|_1\leq \epsilon$. Clearly $K_D (\gamma_m) \ge m$. As every rank $2^m$-dimensional maximally entangled state is a private state, $K_D\ge E_D$. In order to study the question of quantum key repeaters, we introduce the following quantity. For input states $\r_{AC_A}$ between Alice and Charlie and $\tilde{\r}_{C_BB}$ between Charlie and Bob we call
\begin{equation}
K_{A \leftrightarrow C\leftrightarrow B}(\r_{AC_A}\te\tilde{\r}_{C_BB})=\inf_{\e>0}\limsup_{n\to\infty}\sup_{\Lambda_n \LOCC, \gamma_m}\left\{\frac{m}{n}:\tr_{C}\Lambda_n\left(\left(\r_{AC_A}\te\tilde{\r}_{C_BB}\right)^{\te n}\right)\approx_\e\gamma_{ m}\right\}
\end{equation}
the \emph{quantum key repeater rate of $\r$ and $\tilde{\r}$ with respect to arbitrary LOCC operations among Alice, Bob and Charlie}. If we restrict the protocols to one-way communication from Charlie to Alice we write  $K_{A\leftarrow C\leftrightarrow B}$ and if all communication is one-way from Charlie we write $K_{A\leftarrow C\rightarrow B}$.


\pagebreak\section*{Supplementary Note 2}  \label{sub:trace-norm}
\subsection*{Trace Norm Bound}

The distinguishability bound that we present below is based on the notion of distinguishing entangled states from separable states by means of restricted measurements (for example LOCC measurements). Let us briefly describe the derivation of the bound. Consider a state, $\rho_{in}=\rho_{AC_A}\otimes \tilde\rho_{BC_B}$, and suppose $\rho_{in}$ is highly indistinguishable by LOCC operations between $C$ and $AB$  from some triseparable state $\sigma_{in}$. Examples of states $\rho_{in}$ with this property were given in \cite{karol-PhD}: the states are in fact identical private bits $\rho_{AC_A}=\tilde\rho_{BC_B} =\rho$ ($K_D(\rho) = 1$) and  $ \sigma_{in}$ is of the form $\sigma_{AC_A}\otimes \tilde\sigma_{BC_B}$ with $\sigma_{AC_A}=\sigma_{BC_B}$ identical and separable. One may think of them as states that hide entanglement.

Consider now any quantum key repeater protocol $\Lambda $. Since $\Lambda$ is an LOCC operation (between $C$ and $A$ and $B$), its output when acting on $\rho_{in}$ has to be highly indistinguishable by {\it arbitrary} CPTP quantum operations from its output when acting on $\sigma_{in}$. But this means that $\rho_{out}$ and $\sigma_{out}$ are close in trace norm. Since $\sigma_{out}$ is separable this means that $\rho_{out}$ is close to separable and therefore contains almost no key (and is certainly no p-bit). 

To show the above reasoning formally, we first recall the notion of maximal probability 
of discrimination between two states $\rho$ and $\sigma$, using some set $S$ of 
two-outcome POVMs $\{E^{0},E^{1}=\1-E^{0}\}$ \cite{restricted-measurements,karol-PhD}.
By definition we have:
\be 
  p^{S}(\rho,\sigma) = \sup_{\{E^{0},E^{1}\}\in S} \frac12 \Tr E^{0} \rho + \frac12 \Tr E^{1} \sigma.
\ee
In what follows we will consider several sets of operations: $\LOCC$, $\SEP$, $\PPT$ and $\ALL$. 
The set $\ALL$ is the set of all two-outcome POVMs. $\PPT$ consists only of elements that have 
a positive partial transpose and $\SEP$ contains only separable elements, whereas $\LOCC$ are
those POVMs that can be implemented by an LOCC protocol. 
Note that $\LOCC \subset \SEP \subset \PPT \subset \ALL$.

{\lemma For any two states $\rho,\tilde{\rho}$, two separable states $\sigma, \tilde{\sigma}$ and any $\Lambda \in \LOCC(A:C:B)$,
\be 
\|\hat{\rho} -\hat{\sigma}\|_1 \leq \|(\rho_{AC_A}\ot\tilde{\rho}_{BC_B})^{\Gamma} - (\sigma_{AC_A}\ot\tilde{\sigma}_{BC_B})^{\Gamma}\|_1,
\label{eq:gamma-bound}
\ee
where $\hat\rho=\tr_C \Lambda( \rho_{AC_A}\ot\tilde{\rho}_{BC_B})$ and $\hat\sigma=\tr_C\Lambda(\sigma_{AC_A}\ot\tilde{\sigma}_{BC_B})$ are the $AB$ outputs of the protocol. 
\label{lem:main-guess}
}

\proof
Since $\Lambda$ is LOCC, it is a tri-separable map, that is has its Kraus representation $\Lambda(\rho) = \sum_i M_A^i\ot M_B^i\ot M_{C}^i(\rho)M_A^{i\dagger}\ot M_B^{i\dagger}\ot M_{C}^{i\dagger}$. In particular it is separable in the cut $AB:C$, which will be crucial in what follows. Moreover, upon input of any two separable states $\sigma_{AC_A}\ot\sigma_{B_CB}$, the map outputs a state $\rho_{ABC}$ with $\tr_C \rho_{ABC}$ separable. We now prove the following chain of (in)equalities and comment on them below:
\begin{align}
1 + \frac12 \|\hat{\rho} - \hat{\sigma}\|_1  \label{eq:first-helstrom} 
&=2 p^{\ALL}(\hat{\rho},\hat{\sigma})\\
&=\sup_{\{E^{j}\}\in \ALL} [\Tr E^{0}\hat{\rho} + \Tr E^{1} \hat{\sigma}]  \label{eq:def-all}\\
&=\sup_{\{E^{j}_{AB}\}\in\ALL} [\Tr E^{0}_{AB} \Tr_{C}\Lambda(\rho_{AC_A}\ot\tilde{\rho}_{BC_B}) 
            + \Tr E^{1}_{AB} \Tr_{C}\Lambda(\sigma_{AC_A}\ot\tilde{\sigma}_{BC_B})] \label{eq:def-protoc}\\
&=\sup_{\{E^{j}_{AB}\}\in \ALL} [\Tr (E^{0}_{AB}\ot \1_C) \Lambda(\rho_{AC_A}\ot\tilde{\rho}_{BC_B}) 
                               + \Tr (E^{1}_{AB}\ot \1_C) \Lambda(\sigma_{AC_A}\ot\tilde{\sigma}_{BC_B})] \\
&=\sup_{\{E^{j}_{AB}\}\in \ALL} \left[\sum_{j} \Tr (M_A^{j\dagger}\ot M_B^{j\dagger}\ot M_{C}^{j\dagger}(E^{0}_{AB}\ot \1_C)M_A^{j}\ot M_B^{j}\ot M_{C}^{j}(\rho_{AC_A}\ot\tilde{\rho}_{BC_B})) \right. \nonumber \\ 
&\phantom{==:}  
 \left. + \sum_{j} \Tr (M_A^{j\dagger}\ot M_B^{j\dagger}\ot M_{C}^{j\dagger}(E^{1}_{AB}\ot \1_C)M_A^{j}\ot M_B^{j}\ot M_{C}^{j}
(\sigma_{AC_A}\ot\tilde{\sigma}_{BC_B})) \right] \label{eq:key-swap-sep}\\
&\leq 2p^{\SEP(AB:C)}(\rho_{AC_A}\ot\tilde{\rho}_{BC_B},\sigma_{AC_A}\ot\tilde{\sigma}_{BC_B}) \\
&\leq 2p^{\PPT(AB:C)}(\rho_{AC_A}\ot\tilde{\rho}_{BC_B},\sigma_{AC_A}\ot\tilde{\sigma}_{BC_B}) \label{eq:def-sep}\\
&=\sup_{\{F^{j} \geq 0,\sum_{j} F^{j} = \1,(F^{j})^{\Gamma}\geq 0 \}}[\Tr F^{0}(\rho_{AC_A}\ot\tilde{\rho}_{BC_B}) +  \Tr F^{1} (\sigma_{AC_A}\ot\tilde{\sigma}_{BC_B})] \label{eq:gamma-prop-a}\\
&=\sup_{\{F^{j} \geq 0,\sum_{j} F^{j} = \1,(F^{j})^{\Gamma}\geq 0 \}}[\Tr {F^{0}}^\Gamma(\rho_{AC_A}\ot\tilde{\rho}_{BC_B})^\Gamma +  \Tr {F^{1}}^\Gamma (\sigma_{AC_A}\ot\tilde{\sigma}_{BC_B})^\Gamma] \label{eq:gamma-prop}\\
&\leq \sup_{\{\sum_{j} F^{j} = \1,(F^{j})^{\Gamma}\geq 0 \}}[\Tr {F^{0}}^\Gamma(\rho_{AC_A}\ot\tilde{\rho}_{BC_B})^\Gamma + \Tr {F^{1}}^\Gamma (\sigma_{AC_A}\ot\tilde{\sigma}_{BC_B})^\Gamma] \label{eq:all}\\
& = 2 p^{\ALL}((\rho_{AC_A}\ot\tilde{\rho}_{BC_B})^{\Gamma},(\sigma_{AC_A}\ot\tilde{\sigma}_{BC_B})^{\Gamma})  \label{eq:sec-helstrom}\\
& = 1 + \frac12 \|(\rho_{AC_A}\ot\tilde{\rho}_{BC_B})^{\Gamma} -(\sigma_{AC_A}\ot\tilde{\sigma}_{BC_B})^{\Gamma}\|_1.
\end{align}

The first equality is the well known Helstrom formula for optimally distinguishing two quantum states. Subsequently, we simply insert the definitions step by step. Inequality (\ref{eq:key-swap-sep}) follows from the fact that $\Lambda$ is a tri-separable map. In the next inequality we use $\SEP \subset \PPT$. Then we write this explicitly out and partially transpose all the $C$ systems. Then we drop the positivity constraint on the POVM elements and see that the remaining maximisation extends over all POVMs. Using Helstrom once again concludes the calculation. 
\qed

The above lemma shows that the trace norm distance between the output states of any quantum key repeater protocol is upper bounded by the trace norm distance of the partially transposed input states of it. Combining this result with asymptotic continuity of relative entropy of entanglement gives the following theorem:

{\theorem\label{thm:pbit-guess-ex} Consider any two states $\rho,\tilde{\rho}$,
and separable states $\sigma,\tilde{\sigma}$ in $\mathcal{B}(\CC^d\ot \CC^d)$ such that 
$\|\rho^{\Gamma} - \sigma^{\Gamma}\|_1\leq \epsilon$
and $\|\tilde{\rho}^{\Gamma}-\tilde{\sigma}^{\Gamma}\|_1\leq\epsilon$, Then,
if $\mu := \min\{\|\rho^{\Gamma}\|_1,\|\tilde{\rho}^{\Gamma}\|_1\}$ satisfies
$\epsilon' := \epsilon(\mu+1) \leq \frac13$, we have
\be 
  K_\ABC^{\text{single copy}}(\rho\otimes\tilde{\rho}) \leq 4(1+\log d)\epsilon' + 2\eta(\epsilon'),
\ee  
with $\eta(x) = -x \log x$.
Here, $K_\ABC^{\text{single copy}}$ is the quantum key repeater rate when the
repeater is restricted to act on single copies $\rho\otimes\tilde{\rho}$ only.
}

\proof
Let us consider $\|(\rho\otimes\tilde{\rho})^{\Gamma} - (\sigma\otimes\tilde{\sigma})^{\Gamma}\|_1$. By adding and subtracting either $(\rho\otimes\tilde{\sigma})^{\Gamma}$ or $(\sigma\otimes\tilde{\rho})^{\Gamma}$, and by triangle inequality, we obtain
\be 
\|(\rho\otimes\tilde{\rho})^{\Gamma} - (\sigma\otimes\tilde{\sigma})^{\Gamma}\|_1 
     \leq (\min \{ \|\rho^{\Gamma}\|_1,\|\tilde{\rho}^{\Gamma}\|_1\} +1)\epsilon.
\ee
By Lemma \ref{lem:main-guess} and the asymptotic continuity of the relative entropy of entanglement \cite{DonaldH1999} we find
\be 
|E_R(\hat{\rho}) - E_R(\hat{\sigma})|
    \leq 4(1+ \log d) \|\hat{\rho} -\hat{\sigma}\|_1 
             + 2\eta(\|\hat{\rho} -\hat{\sigma}\|_1),
\ee
which, by separability of $\hat{\sigma}$ implies
\be 
E_R(\hat{\rho}) \leq 4  {(1+\log d )}\epsilon' + 2\eta(\epsilon'). 
\ee
Since $K_D\leq E_R$ \cite{pptkey,keyhuge} we have proven the claim.
\qed

\subsection*{Example: p-bit with $X=\text{SWAP}$}
Since the single copy quantum key repeater rate is upper bounded by the general quantum key repeater rate, the example from Supplementary Note 4 can also be used to illustrate the above theorem. We therefore choose to provide an example in this section, which, we believe, is not amenable to the bounds presented elsewhere in this paper. 

We consider $\rho=\tilde{\rho} =\gamma_V$, where $\gamma_V$ is the private state from \cite{pptkey}, shown to be entanglement hiding in \cite{karol-PhD}. It is defined by \eqref{eq:example-gamma} for $X = {\frac{V}{d_s^2}}$ with $V = \sum_{i,j=0}^{d_s-1}|ij\>\<ji|$ the swap operator. Note, that for any private bit described by operator $X$ as in (\ref{eq:example-gamma}), we have $\|\gamma^{\Gamma}\|_1 = 1 + \|X^{\Gamma}\|_1$ (see proof of Theorem 6.5 of \cite{karol-PhD}). Now, following \cite{karol-PhD}, as a state which
is separable and highly indistinguishable from $\gamma_V$, we take $\gamma_V$ dephased on the key part of Alice: $\sigma :=\tilde{\sigma}:={\frac{1}{2}}[\proj{0}\ot \proj{1}\ot \sqrt{XX^{\dagger}} +\proj{1}\ot \proj{0}\ot  \sqrt{X^{\dagger}X}]$. Then $\|\gamma_V^{\Gamma} - \sigma^{\Gamma}\|_1 = \|X^{\Gamma}\|_1$ and $\|X^{\Gamma}\|_1=\|{\frac{V^{\Gamma}}{d_s^2}}\|_1 = \|{\frac{d_s P_+}{d_s^2}}\|_1 = {\frac{1}{d_s}}$ where $P_+ =\frac{1}{d_s}\sum_{i,j=0}^{d_s-1}|ii\>\<jj|$. Thus, 
$\|\gamma_V^{\Gamma} - \sigma^{\Gamma}\|_1 = {\frac{1}{d_s}}$, which for $d_s \geq 7$ by Theorem \ref{thm:pbit-guess-ex} (with $\epsilon' = {\frac{2d_s +1}{d_s^2}}$)  implies that
\be 
  K_\ABC^{\text{single copy}}(\gamma_V\otimes \gamma_V) 
        \leq {\frac{4(2d_s +1)(\log d_s +1)}{d_s^2}} + 2\eta\left(\frac{2d_s +1}{d_s^2}\right).
\ee
Note that the right hand side of the above inequality vanishes with large $d_s$. It cannot be
exactly zero, though, because perfect p-bits always have some non-zero, albeit
sometimes small, distillable entanglement \cite{AH-pditdist}.
This means that $\gamma_V$, although being a private bit ($K_D(\gamma_V) \geq 1$ by definition), 
in fact with $K_D(\gamma_V)=1$ \cite{keyhuge}, cannot be extended by a single copy quantum 
key repeater for large enough $d_s$. 

\pagebreak\section*{Supplementary Note 3} \label{sub:manycopies}

\subsection*{Restricted Relative Entropy Bound}

In this section we derive an asymptotic version of the distinguishability bound, that is, one that upper bounds $K_\ABC$. The quantity which upper bounds the quantum key repeater rate measures the distinguishability of the state to the next separable state in terms of the relative entropy distance of the probability distributions that can be obtained by LOCC. 


Let $\LOCC(A:B)$ be the set of POVMs which can be implemented with local operations and classical communication. We think of an element of this class as the corresponding CPTP map, that is instead of a POVM given by $\{M_i\}$ we consider the CPTP map $M: X \mapsto \sum_i (\Tr M_i X) \proj{i}$. Note that $M(\rho)$ is a probability distribution for $\rho$ a density operator. Our first bound on the quantum key repeater rate is given in terms of the following quantities:
\begin{align}
  D_{C \leftrightarrow AB}(\rho_{AC_A}\otimes \tilde\rho_{C_BB})
    &:=\inf_{\sigma \in \SEP(A:C_A:C_B:B)}\sup_{M \in \LOCC(C:AB)}D(M(\rho\ot \tilde\rho)\|M(\sigma)), \\
  D_{C \rightarrow AB}(\rho_{AC_A}\otimes \tilde\rho_{C_BB})
    &:=\inf_{\sigma \in \SEP(A:C_A:C_B:B)}\sup_{M \in \LOCC(C\rightarrow AB)}D(M(\rho\ot \tilde\rho)\|M(\sigma)).
\end{align}
We denote by $D^\infty$ the regularised versions of the above quantities. 
Note that for trivial $\tilde\rho$, the measures reduce to the measures defined 
in \cite{Piani2009-relent}. Sometimes, we omit the minimisation over separable 
states in which case we write $D_{\CAB}(\rho_{AC_A}\otimes \tilde\rho_{C_BB}\|\sigma_{AC_AC_BB})$.

Before we prove the bound we need an easy lemma that shows that $D_{\ALL}$ 
(as defined by Piani \cite{Piani2009-relent}) is normalised to (at least) 
$m$ on private states $\gamma_m$ \cite{pptkey,keyhuge} containing at least 
$m$ bits of pure privacy.

\begin{lemma} \label{lemma:normalisation} 
For $\tilde\gamma_m\approx_\epsilon \gamma_m$ and $\sigma$ separable we have
\be
  D_{\ALL}(\tilde\gamma_{m}\|\sigma)\geq (1-\epsilon) m-h(\epsilon).
\ee
\end{lemma}

\proof
Recall that $\gamma_m$ is of the form $U P_{m} \otimes \rho_{A'B'} U^\dagger$ for $P_m$ the projector onto the maximally entangled state in dimension $2^m$ on systems $AB$ and $U$ a controlled unitary with control $A$ and target $A'B'$. $\rho_{A'B'}$ is arbitrary. We calculate:

\begin{align}
D_{\ALL}(\tilde\gamma_{m}\|\sigma)&\geq D_{\ALL}(\Tr_{A'B'}U\tilde\gamma_{m}U^\dagger \| \Tr_{A'B'}U\sigma U^\dagger)\\
& =D_{\ALL}(\tilde P_m \| \tilde\sigma) \\
& \geq D(\{\Tr P_m\tilde P_m, \Tr (\1-P_m) \tilde P_m\} \| \{\Tr P_m \tilde\sigma, \Tr (\1 -P_m) \tilde\sigma\}) \\
& \geq  (1-\epsilon) m-h(\epsilon).
\end{align}

The first inequality holds due to monotonicity of $D_{\ALL}$. Note that $\tilde P_m:= \Tr_{A'B'}U\tilde\gamma_{m}U^\dagger$ is a state $\epsilon$-close to $P_m$. We also defined $\tilde\sigma= \Tr_{A'B'}U\sigma U^\dagger$. The second inequality is again an application of monotonicity, this time with the measurement map given by the POVM $\{P_m, \1-P_m\}$. The last inequality follows from the proof of  \cite[Lemma 7]{keyhuge} which says that $\Tr P_m \tilde\sigma \leq 1/2^m$ and $\Tr P_m \tilde P_m \geq 1-\epsilon $, which follows from $\tilde\gamma_m \approx_\epsilon \gamma_m$.
\qed

\medskip
We now come to the main result of this section.
\begin{theorem} \label{theorem:fundamental}
The following inequalities hold for all states $\rho$ and $\tilde\rho$:
\begin{align}
  K_\ABC(\rho_{AC_A}\otimes \tilde\rho_{C_BB})
     &\leq D_\CAB^\infty(\rho_{AC_A}\otimes \tilde\rho_{C_BB}), \\
  K_\ABCr(\rho_{AC_A}\otimes \tilde\rho_{C_BB})
     &\leq D_\CABr^\infty(\rho_{AC_A}\otimes \tilde\rho_{C_BB}).
\end{align}
\end{theorem}
\proof
We will start with proving the first bound. Fix $\epsilon >0$. Then, there is an $n$ and a $\Lambda \in \LOCC(A^n \leftrightarrow C^n \leftrightarrow B^n)$ (in the following we will suppress $n$ if obvious from the context), such that $r \geq K_\ABC(\rho\otimes \tilde\rho)-\epsilon$ and 
$\tilde\gamma:=\Tr_C\Lambda((\rho_{AC_A}\otimes \tilde\rho_{C_BB})^{\otimes n})\approx_\epsilon \gamma_{\lfloor nr \rfloor} $. 
For $\sigma \in \SEP(A:C_A:C_B:B)$ we have
\begin{align}  &\max_{M \in \LOCC(\CAB)}D(M(\rho_{AC_A}^{\otimes n}\otimes \tilde\rho_{C_BB}^{\otimes n})\|M(\sigma_{ACB}))\\
  &\geq   \max_{M \in \LOCC(\CAB)}D(M( \Tr_C\Lambda(\rho_{AC_A}^{\otimes n}\otimes \tilde\rho_{C_BB}^{\otimes n}))\|M(\Tr_C\Lambda(\sigma_{ACB} ))) \\
& = \max_{M\in\ALL(AB)}D(M(\Tr_C\Lambda(\rho_{AC_A}^{\otimes n}\otimes \tilde\rho_{C_BB}^{\otimes n}))\|M(\Tr_C\Lambda(\sigma_{ACB} ))) \\
& = \max_{M \in\ALL(AB)}D(M(\tilde\gamma_{AB})\|M(\tilde\sigma_{AB})).
\end{align}
The first inequality is true as $M\circ \Tr_C\circ \Lambda \in\LOCC(\CAB)$. The first equality follows as the arguments have no system $C$ anymore (or equivalently a one-dimensional system $C$) and since in this case $\LOCC(\CAB)=\ALL(AB)$. In the last equality we have used the definition of $\tilde\gamma$ and introduced $\tilde\sigma:=\Tr_C\Lambda(\sigma )$. Noting that $\tilde\sigma \in \SEP(A:B)$ is separable (since $\Lambda \in\LOCC(\ABC)$ and $\sigma \in \SEP(A:C_A:C_B:B)\subset \SEP(A:C:B)$) and that $\tilde\gamma\approx_\epsilon \gamma_{\lfloor nr\rfloor}$ we have from Lemma \ref{lemma:normalisation}:
\be \max_{M \in\ALL(AB)}D(M(\tilde\gamma_{AB})\|M(\tilde\sigma_{AB}))\geq (1-\epsilon) \lfloor nr \rfloor-h(\epsilon).\ee 
Combining the bounds, minimizing over $\sigma$ and taking the limit $n\rightarrow \infty$ gives
\be D_\CAB^\infty(\rho_{AC_A}\otimes \tilde\rho_{C_BB})\geq (1-\epsilon) r \ee
Since $r \geq K_\ABC(\rho_{AC_A}\otimes \tilde\rho_{C_BB})-\epsilon$ and $\epsilon$ was arbitrary we have proven the first claim. 

The second claim follows by slight modification: restrict $\Lambda$ to be in $\LOCC(A\leftarrow C\rightarrow B)$ and note that $M \circ \Tr_C \circ \Lambda \in\LOCC(C\rightarrow AB)$ and that $\LOCC(C\rightarrow AB)=\ALL(AB)$ for trivial system $C$. Then $K_\ABC$ will turn into $K_\ABCr$ and $D_\CAB$ into $D_\CABr$.
\qed

\subsection*{Properties of the Restricted Relative Entropy Measure}

In this section we present two properties of the distinguishability measure, its invariance under partial transposition of the $C$ system and its LOCC monotonicity. The former provides us with a slightly weaker version of the relative entropy of entanglement bound in Theorem \ref{PPTbound}.

\begin{lemma} \label{lemma:invariance} For all states $\rho$ and $\tilde\rho$,
\begin{align} 
  D_\CAB(\rho_{AC_A}\otimes \tilde\rho_{C_BB}) &= D_\CAB(\rho_{AC_A}^\Gamma \otimes \tilde\rho_{C_BB}^\Gamma), \\
  D_\CABr(\rho_{AC_A}\otimes \tilde\rho_{C_BB})&= D_\CABr(\rho_{AC_A}^\Gamma \otimes \tilde\rho_{C_BB}^\Gamma).
\end{align}
\end{lemma}
\proof
It is sufficient to observe that the sets of measurements which we denote by $\LOCC$ as a placeholder for either $\LOCC(\CAB)$ or $\LOCC(C\rightarrow AB)$ and the set of separable states are invariant under taking partial transpose of systems $C$ (or $AB$):
\begin{align}
\min_{\sigma \in  \SEP(A:C_A:C_B:B)} \max_{M \in\LOCC} &D(M(\rho\otimes \tilde\rho)\|M(\sigma))\\
& =\min_{\sigma \in \SEP(A:C_A:C_B:B)} \max_{M \in\LOCC}D(M^\Gamma(\rho^\Gamma\otimes \tilde\rho^\Gamma)\|M^\Gamma(\sigma^\Gamma))\\
& =\min_{\sigma \in \SEP(A:C_A:C_B:B)}\max_{M \in\LOCC}D(M(\rho^\Gamma\otimes \tilde\rho^\Gamma)\|M(\sigma)).
\end{align}
\qed

By the monotonicity of the relative entropy, we can upper bound $D_{C \leftrightarrow AB}^\infty $ by the relative entropy of entanglement and, using the invariance of  $D_{C \leftrightarrow AB}^\infty $ under partial transpose of the $C$ system (Lemma \ref{lemma:invariance}), obtain

\begin{corollary} \label{cor:boundsRelEnt}
The following inequality holds for all PPT states $\rho_{C_AA}$ and $\tilde{\rho}_{C_BB}$:
\be
  K_\ABC(\rho \otimes \tilde\rho)\leq E^\infty_R(\rho^\Gamma)+E^\infty_R(\tilde\rho^\Gamma).
\ee
\end{corollary}

and thereby almost recover the relative entropy bound from Theorem \ref{PPTbound}.
This lets us also conclude that $D^\infty_{A\leftrightarrow B}(\rho)$, which can similarly be upper bounded by  $E_R(\rho^\Gamma)$, can be made strictly smaller than $K_D(\rho)$: simply take the states from Proposition  \ref{prop:PPT-pbit}. The observation that $D^\infty_{A\leftrightarrow B}$ may be strictly smaller than $K_D$ was first made by Matthias Christandl and Robert Pisarczyk in order to answer a question posed in \cite{LiWinter}.\newline We conclude with proving the monotonicity of the bound.
\begin{lemma}\label{lem:mon}
Let $\Lambda \in \LOCC(C_A \leftrightarrow A)$ and $\Lambda' \in \LOCC(C_A \rightarrow A)$. Then,
\be 
  D_{\CAB}(\rho\ot \tilde\rho)\geq \sum_i p_i D_{\CAB}(\rho_i \ot \tilde\rho),
\ee
and 
\be 
  D_{\CABr}(\rho\ot \tilde\rho)\geq \sum_i p_i' D_{\CABr}(\rho_i' \ot \tilde\rho),
\ee
where $\Lambda(\rho)=\sum_i p_i \proj{i}\ot \rho_i$ and $\Lambda'(\rho)=\sum_i p_i' \proj{i}\ot \rho_i'$. Similar statements hold for $A$ and $B$ exchanged.
\end{lemma}
\proof
We prove the statements for the $\leftrightarrow$ case.
\begin{align}
D_{\CAB}(\rho\ot \tilde\rho)& =\inf_{\sigma \in \SEP(A:C_A:C_B:B)} \max_{M\in\LOCC(\CAB)} D(M(\rho\ot \tilde\rho)\|M(\sigma))\\
& \geq \inf_{\sigma \in \SEP(A:C_A:C_B:B)} \max_{M\in\LOCC(\CAB)} D(M(\Lambda(\rho\ot \tilde\rho))\|M(\Lambda(\sigma)))\\
& = \inf_{\sigma \in \SEP(A:C_A:C_B:B)} \max_{M_i\in\LOCC(\CAB)} \sum_i p_iD(M_i(\rho_i\ot \tilde\rho)\|M_i(\sigma_i))+D(p\|q),
\end{align}
where we used $\Lambda(\sigma)=\sum_i q_i \proj{i}\ot \sigma_i$ and 
without loss of generality $M=\sum_i \proj{i} \ot M_i$.
This is lower bounded by 
\begin{align}
 \inf_{\sigma_i \in \SEP(A:C_A:C_B:B)} \max_{M_i\in\LOCC(\CAB)} \sum_i p_iD(M_i(\rho_i\ot \tilde\rho)\|M_i(\sigma_i)) =\sum_i p_i D_{\CAB}(\rho_i\ot \tilde\rho).
\end{align}
The other cases are similar.
\qed

\subsection*{Squashed Entanglement Bound}

It is the goal of this section to derive a bound on the one-way quantum key repeater rate by the squashed entanglement. For this goal, we need two lemmas in order to prepare for the key lemma, 
Lemma \ref{lemma:KeLi}.
\begin{lemma} \label{lemma:KeLi1}
For any two states $\rho_{ABE}$ and $\sigma_{ABE}$ and for every $M\in\LOCC(A^2\rightarrow B^2)$ with output denoted by $X$ there is a sequence $T_n \in\LOCC(A^n\rightarrow B^n) $ with cq output $X^nB^n$ such that 
\be
   \lim_{n\rightarrow \infty}\frac{1}{n}D(T_n^c(\rho^{\ot n}_{AB})^{\otimes 2}\|T_n^c(\sigma^{\ot n}_{AB})^{\ot 2})=D(M(\rho_{AB}^{\otimes 2})\|M(\sigma_{AB}^{\otimes 2})),
\ee
\be
  \lim_{n\rightarrow \infty}\|T_n^q\ot \id_E(\rho^{\ot n}_{ABE})-\rho^{\ot n}_{BE}\|_1=0,
\ee
where we defined $T_n^q=\Tr_{X^n}\circ T_n$ and $T_n^c=\Tr_{B^n}\circ T_n$.
\end{lemma}
\proof
Apply \cite[Lemma 5]{LiWinter} to the states $\rho\mapsto \rho^{\ot 2}$ and $\sigma\mapsto \sigma^{\ot 2}$. Then manipulate the left hand side of their first equation: First, we use the additivity of the relative entropy
\be D(T_n^c(\rho^{\ot 2n}_{AB})\ot T_n^c(\rho^{\ot 2n}_{AB})\|T_n^c(\sigma^{\ot 2n}_{AB})\ot T_n^c(\sigma^{\ot 2n}_{AB}))=2D(T_n^c(\rho^{\ot 2n}_{AB})\|T_n^c(\sigma^{\ot 2n}_{AB}))\ee in order to conclude 
\begin{align}
\lim_{n\rightarrow \infty}\frac{1}{n}D(T_n^c(\rho^{\ot 2n}_{AB})\|T_n^c(\sigma^{\ot 2n}_{AB}))&=\lim_{n\rightarrow \infty}\frac{1}{2n}D(T_n^c(\rho^{\ot 2n}_{AB})\ot T_n^c(\rho^{\ot 2n}_{AB})\|T_n^c(\sigma^{\ot 2n}_{AB})\ot T_n^c(\sigma^{\ot 2n}_{AB})).
\end{align}
In a next step we restrict the limit to even $n$ (thereby not changing the limiting value) and make the replacement $n \mapsto n/2$ to obtain
\begin{align}
\lim_{n\rightarrow \infty}\frac{1}{n}D(T_{n/2}^c(\rho^{\ot n}_{AB})^{\ot 2}\|T_{n/2}^c(\sigma^{\ot n}_{AB})^{\ot 2}).
\end{align}
Finally, we redefine $T_{n/2} \mapsto T_n$ and obtain the claim. 
\qed

\begin{lemma} \label{lemma:relentlower}
For any tri-tripartite state $\rho$,
\begin{align}
  2E_R^\infty(\rho_{B:AE})\geq D^\infty_{A^2 \rightarrow B^2}(\rho_{AB}^{\ot 2})+2E_R^\infty (\rho_{B:E}).
\end{align}
\end{lemma}
\proof 
For a state $\sigma\in \SEP(B:AE)$,
\begin{align}
nD(\rho^{\otimes 2}_{ABE}\|\sigma_{ABE}^{\ot 2})&=D(\rho^{\otimes 2n}\|\sigma^{\ot 2n})\\
& \geq D(T_n\ot \id_E(\rho^{\ot n})^{\ot 2}\|T_n\ot \id_E(\sigma^{\ot n})^{\ot 2})\\
&= D(T^c_n(\rho^{\ot n})^{\ot 2}\|T^c_n(\sigma^{\ot n})^{\ot 2})+\sum_{ij} p_ip_j D(\rho_i \ot \rho_j\|\sigma_i \ot \sigma_j)\\
& \geq D(T^c_n(\rho^{\ot n})^{\ot 2}\|T^c_n(\sigma^{\ot n})^{\ot 2})+ D(T^q_n\ot \id_E(\rho^{\ot n}) \ot T^q_n\ot \id_E(\rho^{\ot n})\|\tilde\sigma \ot \tilde\sigma)\\
& \geq D(T^c_n(\rho^{\ot n})^{\ot 2}\|T^c_n(\sigma^{\ot n})^{\ot 2})\\
& \quad +\min_{\tilde\sigma \in \SEP(B:E)} D(T^q_n\ot \id_E(\rho^{\ot n}) \ot T^q_n\ot \id_E(\rho^{\ot n})\|\tilde\sigma \ot \tilde\sigma).
\end{align}
The first inequality follows from the monotonicity of the relative entropy under CPTP maps, the following equality is a direct calculation, where the ensemble $\{p_i, \rho_i\}$ ($\{q_i, \sigma_i\}$) is the output of the instrument $T_n\ot \id_E$ when applied to $\rho_{ABE}^{\ot n}$ and $\sigma_{ABE}^{\ot n}$, respectively. The subsequent inequality is due to convexity of the relative entropy, where we defined the state $\tilde\sigma:=T^q_n\ot \id_E(\sigma^{\ot n})$. Since $T^q\ot \id_E\in\LOCC(B:AE)$ and $\sigma\in \SEP(B:AE)$, we find $\tilde\sigma\in \SEP(B:E)$. This explains the last inequality. Using Lemma \ref{lemma:KeLi1}, the asymptotic continuity of the relative entropy of entanglement \cite{DonaldH1999} and taking the limit $n\rightarrow \infty$ proves 
\begin{align}
D(\rho_{ABE}^{\otimes 2}\|\sigma_{ABE}^{\ot 2})\geq D(M(\rho_{AB}^{\ot 2})\|M(\sigma_{AB}^{\ot 2})) +\lim_{n\rightarrow \infty}\frac{1}{n}\min_{\tilde\sigma \in \SEP(B:E)} D(\rho_{BE}^{\otimes n}\otimes \rho_{BE}^{\otimes n}\|\tilde\sigma_{BE}\otimes \tilde\sigma_{BE}) .
 \end{align}
We now maximise this statement over measurements, then minimise over $\sigma$. This proves
\begin{align}
2E_R(\rho_{B:AE}) \geq \inf_\sigma \max_M D(M(\rho_{AB}^{\ot 2})\|M(\sigma^{\ot 2})) +2E_R^\infty (\rho_{B:E}). 
\end{align}
The right hand side is lower bounded by $D_{A^2\rightarrow B^2}(\rho_{AB}\ot \rho_{AB}) +2E_R^\infty (\rho_{B:E})$. 
Regularizing this result we obtain the claimed bound.\qed

\begin{lemma} \label{lemma:KeLi}
\be
  D^\infty_{A^2\rightarrow B^2}(\rho_{AB} \otimes \rho_{AB}) \leq 4E_{sq}(\rho_{AB}).
\ee
\end{lemma}
\proof 
From Lemma \ref{lemma:relentlower} we have 
\begin{align}
 2E_R^\infty(\rho_{B:AE})-2E_R^\infty (\rho_{B:E})\geq  D_{A^2\rightarrow B^2}^\infty(\rho_{AB}^{\ot 2}).
\end{align}
By \cite[Lemma 1]{faithful-squashed} the left hand side is upper bounded by $2I(A:B|E)_\rho$. Minimizing over all extensions of $\rho_{ABE}$ for a fixed $\rho_{AB}$ proves the claim. 
\qed

Combining Lemma \ref{lemma:KeLi} with Theorem \ref{theorem:fundamental} 
and Lemma \ref{lemma:invariance} we get the following bound, which is a weaker version of the squashed entanglement bound in Theorem \ref{PPTbound}

\begin{corollary} \label{cor:boundsquashed}
The following inequality holds for all PPT states $\rho_{C_AA}=\rho_{C_BB}$:
\be
  K_\ABCr(\rho \otimes \rho)\leq  4 E_{sq}(\rho^\Gamma).
\ee
\end{corollary}

\pagebreak\section*{Supplementary Note 4}\label{sub:partialtranspose}
Let us assume that Alice shares a PPT state $\rho$ with Charlie and that Bob shares a PPT state $\tilde\rho$ with Charlie and that they apply an LOCC operation $\Lambda$ among the three of them at the end of which Charlie traces out his part of the system. It is the observation of this section that they obtain the identical output state had they applied the LOCC operation $\Lambda^\Gamma$ (the operation where Charlie's Kraus operators are complex conjugated) to the partially transposed states $\rho^\Gamma$ and $\tilde\rho^\Gamma$ instead. As a consequence, the quantum key repeater rate is invariant under partial transposition:
$K_{A\leftrightarrow C\leftrightarrow B}(\r\te\tilde{\r})
    =K_{A\leftrightarrow C\leftrightarrow B}(\r^\Gamma\te\tilde{\r}^\Gamma)$.
The invariance remains true when restricting partially or fully to one-way communication. In the following, we make this statement precise and use it to find upper bounds. We then give examples illustrating the power of the idea and comparing the obtained bounds.

\subsection*{Bounds by Key, Relative Entropy of Entanglement and Squashed Entanglement}

We start with the above mentioned invariance property.
\begin{lemma}\label{invarianceProperty}
Let $\r$ and $\tilde{\r}$ be PPT. Then
\begin{equation}
  K_{A\leftrightarrow C\leftrightarrow B}(\r\te\tilde{\r})
    =K_{A\leftrightarrow C\leftrightarrow B}(\r^\Gamma\te\tilde{\r}^\Gamma),
\end{equation}
where the transpose is taken w.r.t. Charlie's subsystems.
\end{lemma}
\proof
 Note that every LOCC protocol can be implemented by many rounds of local POVMs and classical communication. If Charlie uses the complex conjugate of all of his Kraus operators $S^{(k)}_C$, we have another valid LOCC protocol. Since 
\begin{align}
&\tr_{C}\left[\left(\ldots \te (S_{C}^{(1)*} \cdots S_{C}^{(r)*})\te \ldots \right)\r^\G_{AC_A}\te\tilde{\r}^\G_{C_BB}\left(\ldots \te ({S_{C}^{(r)*}}^\dagger \cdots {S_{C}^{(1)*}}^\dagger)\te \ldots \right)\right]\\
&=\tr_{C}\left[\left(\ldots \te (S_{C}^{(1)} \cdots S_{C}^{(r)}) \te \ldots \right)\r_{AC_A}\te\tilde{\r}_{C_BB}\left(\ldots \te({S_{C}^{(r)}}^\dagger \cdots {S_{C}^{(1)}}^\dagger)\te \ldots \right)\right],
\end{align}
every protocol applied to copies of $\rho \ot \tilde\rho$ has the same output as when the protocol with complex conjugated Kraus operators is applied to $\rho^\Gamma \ot \tilde\rho^\Gamma$. Consequently, we find
\begin{align}
  K_{\ABC}(\r\te\tilde{\r})&=K_\ABC(\r^\G\te\tilde{\r}^\G).
\end{align}
Recall that this statement only makes sense for PPT states $\rho$ and $\tilde\rho$.
\qed

By the monotonicity of distillable key, we have $K_{\ABC}(\r\te\tilde{\r})\leq K_D(\rho_{AC_A})$. Since the relative entropy of entanglement and squashed entanglement are upper bounds on the key rate \cite{pptkey, Christandl-Schuch-Winter}, that is the right hand side, we obtain the following bounds
\begin{theorem}\label{PPTbound}
Let $\r$ and $\tilde{\r}$ be PPT. Then
\begin{equation}
K_{A\leftrightarrow C\leftrightarrow B}(\r\te\tilde{\r})
  \le\min\left\{ K_D(\rho^\Gamma), K_D(\tilde{\rho}^\Gamma)\right\}
  \le\min\left\{E_R^\infty(\r^\G),E_R^\infty(\tilde{\r}^\G),E_{sq}(\r^\G),E_{sq}(\tilde{\r}^\G)\right\},
\end{equation}
where the transpose is taken w.r.t. Charlie's subsystems.
\end{theorem}

The relative entropy of entanglement \cite{relativeentropy} is given by 
\begin{align}
E_R(\rho)=\inf_{\sigma \in \SEP} D(\rho\|\sigma),
\end{align}
where $\SEP$ denotes the set of separable states.
Since it is subadditive, it upper bounds its regularised version
\begin{align}
E_R^\infty(\rho)=\lim_{n\rightarrow \infty} \frac{1}{n} E_R(\rho^{\otimes n}).
\end{align}
The \emph{squashed entanglement} \cite{Matthias-squashed-ent,T02}
is given by
\begin{equation}
  \label{Esq}
  E_{sq}(\r_{AB}) = \inf_{\r_{ABE}} \frac{1}{2} I(A:B|E)_{\r_{ABE}},
\end{equation}
where $\r_{ABE}$ is an arbitrary extension of $\r_{AB}$.

\subsection*{Example: PPT state close to p-bit}\label{sec:ex-pptpbit}
In the following we exhibit an example, where the right hand sides of our bounds are very small, but where the state itself has a high key rate. The idea here is simple, we find PPT states that have high key but whose partial transpose is close to a separable state \cite{karol-PhD}. More precisely, we present a family of states $\{\rho_{d_s}\}_s$ of increasing dimension which asymptotically reach the gap of 1 between
$K_D(\rho_{d_s})$ and $K_{A\leftrightarrow C\leftrightarrow B}(\rho_{d_s}^{\ot 2})$. Their construction is based on  \cite{smallkey}; there, two private bits were mixed to give a PPT key distillable state. Here we take only one of the p-bits and admix the block-diagonal part of the second one. Alternatively, one may use the family of PPT key distillable states introduced in \cite{pptkey,keyhuge}, but we omit this argument, since it is more involved. 

\begin{proposition}\label{prop:PPT-pbit}
There are PPT states $\rho_{d_s}\in {\cal B}(\CC^{2}\otimes \CC^{2}\otimes \CC^{d_s}\otimes \CC^{d_s})$, obtained by admixing a $p_s$-fraction of a separable state to a p-bit, such that $\rho_{d_s}^\Gamma$ is $p_s$-close to a separable state in trace norm. Furthermore, $p_s = {1\over \sqrt{d_s} +1}$ and $d_s \rightarrow \infty$ for large $d_s$.
\end{proposition}

\proof
Our construction of $\rho_{d_s}$ is based on \cite{smallkey}.
Consider 
\begin{eqnarray}
\rho_{d_s}=\frac{1}{2}\left[ \begin{array}{cccc}
(1-p)\sqrt{X X^{\dagger}} & 0 & 0 & (1-p)X \\
0 & p\sqrt{Y Y^{\dagger}}& 0 & 0 \\
0 & 0 & p \sqrt{Y^{\dagger}Y} & 0\\
(1-p)X^{\dagger} & 0 & 0 & (1-p)\sqrt{X^{\dagger} X}\\
\end{array}
\right],
\label{eq:example}
\end{eqnarray}
with 
\be 
  X=\frac{1}{d_s\sqrt{d_s}} \sum_{i,j=1}^{d_s} u_{ij} |ij\>\<ji|
\label{eq:X}
\ee
and 
\be 
  Y={\sqrt{d_s}}X^{\Gamma}=\frac{1}{d_s} \sum_{i,j=1}^{d_s} u_{ij} |ii\>\<jj|.
\ee
Here, $p_s = {1\over \sqrt{d_s} +1}$ and 
$u_{ij}$ are the matrix elements of some (arbitrary) unitary 
matrix $U$ acting on $\CC^{d_s}$ that satisfies $|u_{ij}|=1/\sqrt{d_s}$ for all $i,j$. 
For example, we may set  $U$ to be quantum Fourier transform 
\be 
  U|k\>=\sum_{j=1}^{d_s} \sqrt{{1\over d_s}}e^{2\pi i j k  /d_s} |j\>.
\ee 

Note that $\rho_{d_s}$ is a mixture of private state (defined by $X$) 
with probability $1-p$ and a with separable state 
${1\over 2}[\proj{0}\otimes \proj{1}\otimes \sqrt{YY^{\dagger}} + \proj{1}\otimes \proj{0}\otimes \sqrt{Y^{\dagger}Y}]$ 
with probability $p$. It is easy to see that the state is PPT, 
as $(1-p)X^{\Gamma} = p Y$. So after partial transposition of $BB'$:
\begin{eqnarray}
\rho_{d_s}^{\Gamma}=\frac{1}{2}\left[ \begin{array}{cccc}
(1-p)\sqrt{X X^{\dagger}} & 0 & 0 & 0 \\
0 & p\sqrt{Y Y^{\dagger}}& p Y & 0 \\
0 & p Y^{\dagger} & p \sqrt{Y^{\dagger}Y} & 0\\
0 & 0 & 0 & (1-p)\sqrt{X^{\dagger} X}\\
\end{array}
\right],
\label{eq:ppt}
\end{eqnarray}
which is evidently non-negative, as $\sqrt{XX^{\dagger}}$ and $\sqrt{X^{\dagger}X}$ 
are non-negative by definition, and the middle block is 
(up to normalisation factor $p$) a private bit defined by operator $Y$ \cite{keyhuge}.

Consider now the state $\rho_{d_s}$ dephased on the first qubit of Alice's system 
(this state is also known as ``key attacked state''). It reads:
\begin{eqnarray}
\sigma_{d_s}=\frac{1}{2}\left[ \begin{array}{cccc}
(1-p)\sqrt{X X^{\dagger}} & 0 & 0 & 0 \\
0 & p\sqrt{Y Y^{\dagger}}& 0 & 0 \\
0 & 0 & p \sqrt{Y^{\dagger}Y} & 0\\
0 & 0 & 0 & (1-p)\sqrt{X^{\dagger} X}\\
\end{array}
\right],
\label{eq:dephased}
\end{eqnarray}
and is clearly separable. It is easy to see that
\be 
  \|\rho_{d_s}^{\Gamma} - \sigma_{d_s}^{\Gamma}\|_1
     = \|(1-p)X^{\Gamma}\|_1=\|pY\|_1 = p = \frac{1}{\sqrt{d_s} + 1}.
\label{eq:small-dist}
\ee
This concludes the proof.
\qed

Since the states $\rho_s$ are obtained by admixing a small fraction of a separable state to a p-bit, the key rate of the state is high: Alice and Bob's mutual information in fact equals $1-h(p_s)$ and the quantum mutual information of Alice and Eve is bounded by $h(p_s)$. Hence, by \cite{DevetakWinter-hash}, $K(\rho)\geq 1-2 h(p_s)$. On the other hand, $\rho^\Gamma$ is almost separable which implies that $K(\rho^\Gamma)$, $E_R(\rho^\Gamma)$ and $E_{sq}(\rho^\Gamma)$ are small. A particularly good bound is obtained with help of the following lemma.

\begin{lemma} \label{lemma:guessbound}
Let $\rho_{ABA'B'} \in {\cal B}(\CC^{2}\otimes \CC^{2}\otimes \CC^{d}\otimes \CC^{d})$ be 
a $\PPT(AA':BB')$ state and assume that its key attacked version
$\sigma_{ABA'B'}=\sum_i (\ket{i}\bra{i}_A \otimes {\normalfont \1}) 
                                                     \rho (\ket{i}\bra{i}_A\otimes {\normalfont \1})$ 
is separable. Then if $\epsilon= \|\rho^\Gamma - \sigma^\Gamma\|_1 < {1\over 3}$, we have
\be 
  E^{\infty}_R(\rho^\Gamma) \leq 2\epsilon\log 2d + \eta(\epsilon),
\ee
where $\eta(\epsilon) = -\epsilon \log \epsilon$.
\end{lemma}

\proof
We start by noting that $\sigma$ and hence $\sigma^\Gamma$ are separable, therefore we have 
\be E_R^{\infty}(\rho^{\Gamma})\leq E_R(\rho^\Gamma)\leq D(\rho^\Gamma\|\sigma^\Gamma)\ee
We write out the right hand side
\be D(\rho^\Gamma\|\sigma^\Gamma)= \Tr \rho^\Gamma \log \rho^\Gamma-\Tr \rho^\Gamma \log \sigma^\Gamma.\ee
and find, since $\Tr \rho^\Gamma \log \sigma^\Gamma=\Tr \sigma^\Gamma \log \sigma^\Gamma$ (due to the fact that $\sigma$ is block diagonal) that 
\be D(\rho^\Gamma\|\sigma^\Gamma)=H(\sigma^\Gamma)-H(\rho^\Gamma).\ee
An application of Fannes' inequality \cite{Fannes1973} gives the result.
\qed

\begin{theorem}\label{thm:key2} 
There are PPT states $\rho_{d_s}\in {\cal B}(\CC^{2}\otimes \CC^{2}\otimes \CC^{d_s}\otimes \CC^{d_s})$, satisfying 
$K_D(\rho_{d_s}) = 1 - 2h(p_s)$ with $p = \frac{1}{\sqrt{d_s}+1}$ and $h$ the binary Shannon entropy, such that
$K_\ABC(\rho_{d_s}\otimes \rho_{d_s})\leq 2{p} \log(2d_s) + \eta(p)$ where $\eta(p)=-p\log p$. In summary, there exist states with 
\be
  1 \approx K_D(\rho) > K_{A\leftrightarrow C \leftrightarrow B}(\rho \otimes \rho) \approx 0.
\ee
\end{theorem}

\subsection*{Comparison of the Bounds: Werner States}

In the following we show that the bound by the squashed entanglement can be smaller than the one by the relative entropy of entanglement. Recall that it was previously known that squashed entanglement of the antisymmetric Werner state is smaller than its relative entropy of entanglement \cite{CSW09, Christandl-Schuch-Winter}. Since the antisymmetric Werner state is not PPT, however, this example does not apply directly to our situation. Using a related PPT state from \cite{AEJPVM2001}, we are able to obtain our goal. We leave open the question of whether the relative entropy of entanglement can be smaller than squashed entanglement. This, however, seems very plausible, as squashed entanglement is lockable \cite{ChristandlW-lock}, and the relative entropy is not \cite{lock-ent}. The challenge therefore remains to show locking of squashed entanglement for a PPT state.

Let $\tau_\pm$ be the symmetric and antisymmetric Werner state. In \cite{AEJPVM2001} it is shown that 
\be \rho^n:=w\tau_-^{\ot n}+ (1-w) \tau^{\ot n}\ee
is PPT for $w=1/(1+z^n)$ for $z=(d+2)/d$, $p=(d+1)/(d+2)$ and $\tau:=(1-p) \tau_-+p\tau_+$. Note that 
\be 
  E_{sq}(\rho^n)\leq n E_{sq}(\tau_-), 
\ee
since $\tau$ is separable. By a result of \cite{Christandl-Schuch-Winter}, $E_{sq}(\tau_-)\leq O(1/d)$ hence we find
\be
  E_{sq}(\rho^n)\leq O(n/d).
\ee
Let us now derive a lower bound on the regularised relative entropy of this state.
Since the relative entropy is not lockable we find
\begin{align}
  E_R((\rho^{ n})^{\otimes k})
     &\geq \sum_j {k \choose j} w^j (1-w)^{k-j} E_R(\tau_-^{\ot j n}\ot \tau^{\ot (k-j)n})-kh(w)\\
     &=    \sum_j {k \choose j} w^j (1-w)^{k-j} E_R(\tau_-^{\ot j n})-kh(w)\\
     &\approx  E_R(\tau_-^{\ot wkn})-kh(w),
\end{align}
where we used the separability of $\tau$ in the first equality and the law of large numbers in the second. Taking the large $k$ limit we find
\be
  E_R^\infty(\rho^{ n}) \geq wnE_R^\infty(\tau_-)-h(w).
\ee
By \cite{Christandl-Schuch-Winter}, $E_R^\infty(\tau_-)$ is 
lower bounded by a constant independent of $d$. 
Setting $n=O(d)$ we find $w=O(1)$ (which can be made arbitrarily small) 
and hence $E_R^\infty(\rho^{ n}) \geq O(n)$. From the bound above $E_{sq}(\rho^n)\leq O(1)$.
Hence there are PPT states $\hat\rho$ for which
\be
  E_{sq}(\hat\rho)\ll E_R^\infty(\hat\rho).
\ee
Since $\rho:=\hat\rho^\Gamma$ is again a PPT state we also find that there are PPT states $\rho$ for which
\be
  E_{sq}(\rho^\Gamma)\ll E_R^\infty(\rho^\Gamma).
\ee
This shows that the squashed entanglement bound may be stronger than the regularised relative entropy bound.



\pagebreak\section*{Supplementary Note 5} \label{sub:stefan-andreas}


\subsection*{Entanglement Distillation and Cost Bound}
We will now present an upper bound on the quantum key repeater rate that depends on the distillable entanglement of the input state.
\begin{theorem}\label{main_theo}
For input states $\rho_{AC_A}$ and $\tilde{\rho}_{C_BB}$ it holds
\begin{align}
K_\ABCrr(\rho_{AC_A}\te\tilde{\rho}_{C_BB})& \le\frac{1}{2}E_D(\tilde{\rho}_{C_BB})+\frac{1}{2}E_C(\rho_{AC_A}),\\
K_\ABCr(\rho_{AC_A}\te\tilde{\rho}_{C_BB}) &\le\frac{1}{2}E_D^{C_B\to B}(\tilde{\rho}_{C_BB})+\frac{1}{2}E_C(\rho_{AC_A}),
\end{align}
where $E_D^{C_B\to B}$ denotes the \textit{one-way distillable entanglement}. In case of PPT states, we may also transpose the states on the $C$ system.
\end{theorem}
Our result implies that if one of the input states is bound entangled or has small distillable entanglement, the other state has to 'compensate' this lack of distillability by its entanglement cost. Before proving Theorem \ref{main_theo}, we consider the \emph{classical squashed entanglement} \cite{T02}, denoted by $E_{sq,c}$, a variant of the squashed entanglement where the extensions are restricted to being classical, that is 
$\r_{ABE}=\sum_ip_i\r^{(i)}_{AB}\te\pro{i}_{E}$.
If we further restrict ourselves to  
$\r_{ABE}=\sum_ip_i\pro{\Psi^{(i)}}_{AB}\te\pro{i}_{E}$, that is~pure states
$\rho_i = \pro{\Psi^{(i)}}$, we get the \emph{entanglement of formation} 
 \cite{T02}. Clearly, $E_{sq} \le E_{sq,c} \le E_F$, and all inequalities
can be strict, for example for the antisymmetric state~\cite{CSW09,Christandl-Schuch-Winter}.
Furthermore, in \cite{MC-phd,CSW09,Christandl-Schuch-Winter} it was shown that $K_D\le E_{sq}$. The proof of 
 Theorem \ref{main_theo} is based on the following lemmas. First, a small technical observation:
\begin{lemma}
  \label{lemma1}
  For any bipartite state $\r_{AB}$ it holds $E_D(\r_{AB})\ge E_D^{B\to A}(\r_{AB})\ge2E_{sq,c}(\r_{AB})-H(B)_\r$.
\end{lemma}
\proof
Using the definition of the classical squashed entanglement and the 
hashing inequality \cite{DevetakWinter-hash}, we have 
$2E_{sq,c}(\r_{AB})\le I(A:B)_\r=H(B)_\r-H(B|A)_\r\le H(B)_\r+E_D^{B\to A}(\r_{AB})$.
\qed
Lemma \ref{lemma1} gives us the following upper bound on the classical squashed entanglement of $\tau$:

\begin{lemma}\label{th1}
For $\LOCC(\ABCrr)$ protocols resulting in $\tau_{A'B'}$ there holds
\be  E_{sq,c}(\tau_{A'B'})\le\frac{1}{2}E_D(\tilde{\rho}_{C_BB})+ \frac{1}{2}E_F(\rho_{AC_A}).
\ee

\end{lemma}

\proof
For any $\LOCC(\ABCrr)$ protocol there exists a two step protocol of the following form that results in the same state: \textit{First}, Charlie and Bob perform an LOCC operation $\Lambda$ on their subsystems after which Charlies system is discarded. As part of $\Lambda$, any classical message intended for Alice is stored at Bobs site, for now. This results in a state $\sigma_{AB'}=\tr_C\left[\1_A\te\Lambda_{CB}\left(\rho_{AC_A}\te\tilde{\rho}_{C_BB}\right)\right]$, where Alices message is contained in the $B'$ subsystem. In a \textit{second} step, Bob sends the classical message to Alice who then performs a local operation depending on the message. This results in state $\tau_{A'B'}$.\newline

Let $\{q_j,\pro{\Ps_j}_{AC_A}\}$ be an ensemble such that $\rho_{AC_A}=\sum_jq_j\pro{\Ps_j}_{AC_A}$ and $E_F(\rho_{AC_A})=\sum_jq_j H(A)_{\pro{\Ps_j}}$. For every $j$, applying the first step of the protocol to $\pro{\Ps_j}_{AC_A}\te\tilde{\rho}_{C_BB}$ alone results in a state 
$\sigma^{(j)}_{AB'}=\tr_C\left[\1_A\te\Lambda_{CB}\left(\pro{\Ps_j}_{AC_A}\te\tilde{\rho}_{C_BB}\right)\right]$. By linearity we have $\s_{AB'}=\sum_{j}q_j\s^{(j)}_{AB'}$. By Lemma \ref{lemma1}, it holds 
\be 
E_D(\s^{(j)}_{AB'})\ge2E_{sq,c}(\s^{(j)}_{AB'})-H(A)_{\s^{(j)}}=2E_{sq,c}(\s^{(j)}_{AB'})-H(A)_{\pro{\Psi_j}},
\ee
where I have used the fact that the $A$ subsystem remains untouched in the first step. Applying the convex sum results in
\be 
\sum_{j}q_jE_D(\s^{(j)}_{AB'})\ge\sum_{j}q_j2E_{sq,c}(\s^{(j)}_{AB'})-E_F(\rho_{AC_A}).
\ee

As the second step of the protocol is LOCC, using the convexity and LOCC monotonicity of the classical squashed entanglement \cite{YHHHOS09}, we obtain $\sum_{j}q_j E_{sq,c}(\s^{(j)}_{AB'})\ge E_{sq,c}(\tau_{A'B'})$. In order to get rid of the convex sum in front of $E_D$, one can apply its LOCC monotonicity in a scenario where Alice and Charlie are sharing a lab. Namely, we need an $LOCC(AC\leftrightarrow B)$ protocol, transferring $\tilde{\rho}_{C_BB}$ into the ensemble $\{q_j,\s^{(j)}_{AB'}\}$. Such a protocol exists: If Alice and Charlie share a lab they will be able to locally create the ensemble $\{q_j,\pro{\Ps_j}\}$. Then all that is left to do is to apply the first part of the swapping protocol. By the LOCC monotonicity of $E_D$ it holds $E_D(\tilde{\rho}_{C_BB})\ge\sum_{j}q_j E_D(\s^{(j)}_{AB'})$, finishing the proof.
\qed
Similarly, we can show the following

\begin{lemma}\label{th2}
For $\LOCC(\ABCr)$ protocols resulting in $\tau_{A'B'}$ there holds
\be  E_{sq,c}(\tau_{A'B'})\le\frac{1}{2}E_D^{C_B\to B}(\tilde{\rho}_{C_BB})+ \frac{1}{2}E_F(\rho_{AC_A}).
\ee

\end{lemma}

\proof
For any $\LOCC(\ABCr)$ protocol there exists a two step protocol of the following form that results in the same state: \textit{First}, Charlie and Bob perform an LOCC($C\to B$) operation $\Lambda$ on their subsystems after which Charlies system is discarded. As part of $\Lambda$, any classical message intended for Alice is stored at Bobs site, for now. This results in a state $\sigma_{AB'}=\tr_C\left[\1_A\te\Lambda_{CB}\left(\rho_{AC_A}\te\tilde{\rho}_{C_BB}\right)\right]$, where Alices message is contained in the $B'$ subsystem. In a \textit{second} step, Bob sends the classical message to Alice who then performs a local operation depending on the message. This results in state $\tau_{A'B'}$.\newline

Let $\{q_j,\pro{\Ps_j}_{AC_A}\}$ be an ensemble such that $\rho_{AC_A}=\sum_jq_j\pro{\Ps_j}_{AC_A}$ and $E_F(\rho_{AC_A})=\sum_jq_j H(A)_{\pro{\Ps_j}}$. For every $j$, applying the first step of the protocol to $\pro{\Ps_j}_{AC_A}\te\tilde{\rho}_{C_BB}$ alone results in a state 
$\sigma^{(j)}_{AB'}=\tr_C\left[\1_A\te\Lambda_{CB}\left(\pro{\Ps_j}_{AC_A}\te\tilde{\rho}_{C_BB}\right)\right]$. By linearity we have $\s_{AB'}=\sum_{j}q_j\s^{(j)}_{AB'}$. By Lemma \ref{lemma1}, it holds 
\be 
E_D^{A\to B'}(\s^{(j)}_{AB'})\ge2E_{sq,c}(\s^{(j)}_{AB'})-H(A)_{\s^{(j)}}=2E_{sq,c}(\s^{(j)}_{AB'})-H(A)_{\pro{\Psi_j}},
\ee
where I have used the fact that the $A$ subsystem remains untouched in the first step. Applying the convex sum results in
\be 
\sum_{j}q_jE_D^{A\to B'}(\s^{(j)}_{AB'})\ge\sum_{j}q_j2E_{sq,c}(\s^{(j)}_{AB'})-E_F(\rho_{AC_A}).
\ee

As the second step of the protocol is LOCC, using the convexity and LOCC monotonicity of the classical squashed entanglement \cite{YHHHOS09}, we obtain $\sum_{j}q_j E_{sq,c}(\s^{(j)}_{AB'})\ge E_{sq,c}(\tau_{A'B'})$. In order to get rid of the convex sum in front of the one-way distillable entanglement, one can apply its LOCC monotonicity in a scenario where Alice and Charlie are sharing a lab. Namely, we need an $LOCC(AC\to B)$ protocol, transferring $\tilde{\rho}_{C_BB}$ into the ensemble $\{q_j,\s^{(j)}_{AB'}\}$. Such a protocol exists: If Alice and Charlie share a lab they will be able to locally create the ensemble $\{q_j,\pro{\Ps_j}\}$. Then all that is left to do is to apply the first part of the swapping protocol. By the one-way LOCC monotonicity of the one-way distillable entanglement it holds $E_D^{C_B\to B}(\tilde{\rho}_{C_BB})\ge\sum_{j}q_j E_D^{A\to B'}(\s^{(j)}_{AB'})$, finishing the proof.
\qed

We are now ready to prove Theorem \ref{main_theo}.

\proof\textbf{of Theorem \ref{main_theo}\ }
Let $\cM$ be the class of allowed LOCC protocols and let $\e>0$. Then there exists $n$ and an $\cM$-protocol $\Lambda^{\cM}$ such that $\tr_{C}\Lambda^{\cM}\left(\left(\rho\te\tilde{\rho}\right)^{\te n}\right)\approx_\e\g_{\left\lfloor nr\right\rfloor}$ and $r\ge K_\cM(\rho\te\tilde{\rho})-\e$. Hence, using the fact that $E_{sq}(\g_m)\ge m$ for any $\g_m$ \cite{CSW09}, as well as the LOCC monotonicity and asymptotic continuity of $E_{sq}$, it holds
\begin{equation}
nK_\cM(\rho\te\tilde{\rho})\le nr+n\e\le E_{sq}(\g_{\left\lfloor nr\right\rfloor})+n\e\le E_{sq}\left(\tr_C\Lambda^{\cM}\left((\rho\te\tilde{\rho})^{\te n}\right)\right)+\text{const}\e\log(\text{dim}^n_{A'B'})+f(\e)+n\e,
\end{equation}
where $f(\e)\to0$ as $\e\to0$. By Lemma \ref{th1} and \ref{th2} for respective classes $\cM$ and the fact that $E_{sq}\le E_{sq,c}$, it holds
\begin{equation}
E_{sq}\left(\tr_C\Lambda^{A\leftarrow C\leftrightarrow B}\left((\rho\te\tilde{\rho})^{\te n}\right)\right)\le\frac{1}{2}E_D(\tilde{\rho}^{\te n})+\frac{1}{2}E_F(\rho^{\te n})
\end{equation}
and
\begin{equation}
E_{sq}\left(\tr_C\Lambda^{A\leftarrow C\rightarrow B}\left((\rho\te\tilde{\rho})^{\te n}\right)\right)\le\frac{1}{2}E_D^{C_B\to B}(\tilde{\rho}^{\te n})+\frac{1}{2}E_F(\rho^{\te n}).
\end{equation}
Let us now divide by $n$ and let $\e\to0$ and $n\to\infty$. Our bounds then follow from the extensitivity of $E_D$ and the fact that the regularised entanglement of formation equals the entanglement cost. If $\rho$ and $\tilde{\rho}$ are PPT, it can be shown analogously to Lemma \ref{invarianceProperty} that $K_{A\leftarrow C\leftrightarrow B}(\rho\te\tilde{\rho})=K_{A\leftarrow C\leftrightarrow B}(\rho^\Gamma\te\tilde{\rho}^\Gamma)$ and $K_{A\leftarrow C\rightarrow B}(\rho\te\tilde{\rho})=K_{A\leftarrow C\rightarrow B}(\rho^\Gamma\te\tilde{\rho}^\Gamma)$, hence we can also partially transpose $\rho$ and $\tilde{\rho}$.
\qed



\subsection*{Example: PPT invariant approximate p-bit (based on data hiding states)}

Note that, even though the results in Section \ref{sub:manycopies} may be computed for states without the use of the partial transpose, all examples were in fact computed using that idea. Therefore, until now, we have not been able to demonstrate a nontrivial bound for states that are invariant under the partial transpose operation. It is the goal of this section to demonstrate such an example by help of Theorem \ref{main_theo}.

In order to do so, we choose a family of states $\rho_m$ and based on this, consider states of the form $\tilde{\rho}_m := \rho_m\otimes \rho_m^{\Gamma}$. Note that $ \tilde{\rho}_m$ is locally equivalent (by bilocal swap) to its partial transposition. The bounds on using the partial transposition which we presented earlier do therefore not give any interesting bounds in this situation. As we show below, however, for our choice of $\tilde{\rho}_m$ we find $E_D( \tilde{\rho}_m)=0$ and $E_C( \tilde{\rho}_m) \lesssim 1$. 
Inserting this into Theorem \ref{main_theo}, we find
\be
  K_{A\leftarrow C \leftrightarrow B}( \tilde{\rho}_m \ot  \tilde{\rho}_m)\lesssim \frac{1}{2},
\ee
which is significantly smaller than $K_D( \tilde{\rho}_m) \gtrsim 1$ (see below).

In order to construct $\rho_m$, we consider a family of states on 
$B\left(\CC^2\ot\CC^2\ot(\CC^{d^k}\ot\CC^{d^k})^{\ot m}\right)$ given in \cite{pptkey}:
\be 
\hat{\rho}_{p,d,k,m} ={1\over N_{m}}\left[\begin{array}{cccc}
[p({\tau_1+\tau_2\over 2})]^{\ot m} &0&0&[p({\tau_1-\tau_2\over 2})]^{\ot m} \\
0& [({1\over 2}-p)\tau_2]^{\ot m}&0&0 \\
0&0&[({1\over 2}-p)\tau_2]^{\ot m}& 0\\
{[p({\tau_1-\tau_2\over 2})]}^{\ot m}&0&0&{[p({\tau_1+\tau_2\over 2})]}^{\ot m}\\
\end{array}
\right],
\label{eq:rec-state-presented}
\ee
where $N_m = 2(p^m)+2({1\over 2} -p)^m$, $\tau_1 = ({\rho_a + \rho_s\over 2})^{\ot k}$ and $\tau_2=(\rho_s)^{\ot k}$, while $\rho_s$ and $\rho_a$ are the $d$-dimensional symmetric and antisymmetric Werner state, respectively.

The state $\hat{\rho}_{p,d,k,m}$ is PPT iff $p \leq \frac13$ and 
$\frac{1-p}{p} \geq (\frac{d}{d-1})^k$ \cite{pptkey}. 
We satisfy this condition by setting $p=\frac13$, 
$d =m^2$ and $k = m$, as then $({d\over d-1})^k < 2$ for $m \geq 2$. 
Then we define
\be 
  \rho_m := \hat{\rho}_{1/3, m^2,m,m},
  \label{eq:rhom}
\ee
with $m \geq 2$. Since also $\tilde{\rho}_m$ is PPT, it is bound 
entangled and we find $E_D( \tilde{\rho}_m)=0$. 
The following lemma assures us of the fact that entanglement of 
formation of $\tilde{\rho}_m$ is bounded by approximately one.
 
{\lemma $\tilde{\rho}_m = \rho_m\otimes \rho_m^{\Gamma}$ for $\rho_m$ defined in eq.~(\ref{eq:rhom}) 
satisfies $E_C(\tilde{\rho}_m)\leq E_F(\tilde{\rho}_m) \leq 1+ {2m^2 \log(2m) \over 2^{m}+1} $. 
Note that this bound is approximately equal to one for large $m$.
\label{lem:eform-bounded}
}

{\proof 
Observe first that $E_F(\tilde{\rho}_m) \leq E_F(\rho_m)+ E_F(\rho^{\Gamma}_m)$ due to the subadditivity of $E_F$.
We show now, that $E_F(\rho_m)\leq 1$. Indeed,
observe that (for $x=\frac{(1/2-p)^m}{N_m}$)
\ben
 \rho_m = (1 - 2x) \left[ \frac12 |\psi_+\>\<\psi_+|\otimes S_\text{even} + \frac12 |\psi_-\>\<\psi_-|\otimes S_{\text{odd}} \right] + \nonumber\\
 2 x \left[ \frac12 |01\>\<01|\otimes \tau_2^{\otimes m} + \frac12 |10\>\<10|\otimes \tau_2^{\ot m} \right],
\een 
where $S_{\text{even}}$ is a uniform mixture (with probability $2^{-(m-1)}$) of all states 
$\tau_{i_1}\otimes \cdots\otimes \tau_{i_m}$ such that $2$ occurs even number 
of times in string $(i_1,\ldots,i_m)$, and $S_{\text{odd}}$ is defined analogously, but with number of $2$ being odd,  $|\psi_\pm\> = {1\over \sqrt{2}}(|00\> \pm |11\>)$. It is clear from the above formula,
that the state $\rho_m$ can be created from 2-qubit maximally entangled state appropriately correlated to the sequences of length $m$ of separable hiding states $\tau_i$, and mixed with probability $2x$ with a separable state ${1\over 2}(|01\>\<01|\otimes \tau_2^{\otimes m} + |10\>\<10|\otimes \tau_2^{\ot m})$.

We now bound $E_F(\rho^{\Gamma}_m)$ from above. Note that 
\be 
\rho^{\Gamma}_m={1\over N_{m}}\left[\begin{array}{cccc}
[p({\tau_1+\tau_2\over 2})^{\Gamma}]^{\ot m} &0&0&0 \\
0& [({1\over 2}-p)\tau_2^{\Gamma}]^{\ot m}&[p({\tau_1-\tau_2\over 2})^{\Gamma}]^{\ot m}&0 \\
0&[p({\tau_1-\tau_2\over 2})^{\Gamma}]^{\ot m}&[({1\over 2}-p)\tau_2^{\Gamma}]^{\ot m}& 0\\
0&0&0&{[p({\tau_1+\tau_2\over 2})^{\Gamma}]}^{\ot m}\\
\end{array}
\right],
\ee
Observe, that $[({\tau_1+\tau_2\over 2})^{\Gamma}]$ is a separable state, and, therefore, by the convexity of entanglement of formation, $E_F(\rho_m^\Gamma) \leq 2x E_F(\rho'_m)$ where the state $\rho'_m$ is formed by middle block of the above matrix:
\be 
\rho'_m ={1\over 2({1\over 2}-p)^m}\left[\begin{array}{cccc}
0 &0&0&0 \\
0& [({1\over 2}-p)\tau_2^{\Gamma}]^{\ot m}&[p({\tau_1-\tau_2\over 2})^{\Gamma}]^{\ot m}&0 \\
0&[p({\tau_1-\tau_2\over 2})^{\Gamma}]^{\ot m}&[({1\over 2}-p)\tau_2^{\Gamma}]^{\ot m}& 0\\
0&0&0&0\\
\end{array}
\right].
\ee
Since $x \leq {1\over 2^m}$, we can safely bound $E_F(\rho'_m)$ by the logarithm of the local dimension of $\rho'_m$, which equals $2m^{2m^2}$: 
\be 
E_F(\rho^{\Gamma}_m) \leq 2x \times 2 m^2 \log(2m).
\ee
The assertion follows by inserting $p=1/3$ and observing that the entanglement cost is upper bounded by the entanglement of formation. \qed
}

In the following we show that $K_D(\tilde\rho_m) \gtrsim1$ in the limit of large $m$. We start by noting that $K_D(\tilde\rho_m)\geq K_D(\rho_m)$ and that it therefore suffices to lower bound $K_D(\rho_m)$. We first apply a privacy squeezing operation
to $\rho_m$, which gives $\rho_m^{ps}$ \cite{keyhuge}. Note, that this operation on $\rho_m$ amounts to the replacement of the blocks of the matrix given in eq. (\ref{eq:rec-state-presented}) by their respective 
trace norms. In turn, the $\rho_m^{ps}$ is a 2-qubit state described by the matrix:
\be 
\left[\begin{array}{cccc}
a &0&0&b \\
0& x&0&0 \\
0&0&x& 0\\
b&0&0&a\\
\end{array}
\right],
\ee
where $a = {p^m\over N_m}$, $x = {(1/2 - p)^m\over N_m}$ and (by eq. 141 of \cite{keyhuge}) $b = {(p(1 - 2^{-m}))^m\over N_m}$. Now, using the fact that the distillable key of $\rho_m$ is lower bounded by the Devetak-Winter quantity of a ccq state of the $\rho_m^{ps}$ (see Corollary 4.26 of \cite{karol-PhD}), we observe that:  
\be 
K_D(\rho_m) \geq 1 - H(a+b,a-b,x,x),
\label{eq:kd-ps-bound}
\ee
where $H$ is the Shannon entropy. This is what we aimed to prove, as in the limit of large $m$ the above 
considered distribution approaches $(1,0,0,0)$ for our choice of $p$.
\qed

\subsection*{Private states with bounded key repeater rate}

In this section we provide a family of private bits $\gamma_{m}$, such that $K_{A\leftarrow C \leftrightarrow B}$ approaches ${1\over 2}$ 
for large $m$. In \cite{keyhuge}, it is proven that provided a certain submatrix of a state $\rho \in B(C^2\otimes C^2\otimes C^d\otimes C^d$) has large enough trace norm, there exists a private bit $\gamma$ which is close to $\rho$ in trace norm. Moreover, the construction of $\gamma$ is explicit. We choose $\rho=\rho_m$, given in \eqref{eq:rhom}, as it has $E_D(\rho_m) = 0$ and $E_F(\rho_m)\leq 1$. Using entanglement theory, we show, that the constructed $\gamma_m$ satisfies $E_D(\gamma_m)\approx 0$ and $E_F(\gamma_m)\approx 1$ for 
large enough $m$. Finally we use theorem \ref{main_theo}, which under these conditions proves $K_{A\leftarrow C \leftrightarrow B}(\gamma_m) \approx {1\over 2}$ for large $m$.

We start by recalling the following result.
{\proposition \cite{keyhuge}
If the state $\sigma_{ABA'B'} \in {\cal B}({\cal C}^2\ot{\cal C}^2 \ot {\cal C}^d\ot{\cal C}^{d'}) $ with a form $\sigma_{ABA'B'} = \sum_{ijkl=0}^{1} |ij\>\<kl|_{AB}\ot A_{ijkl}$  fulfills 
$|| A_{0011}||_1 > {1\over 2}-\epsilon$ for some $0< \epsilon <{1\over 8 e^2}$, then there exists private bit $\gamma$ such, that  
\be
|| \sigma_{ABA'B'} - \gamma_{ABA'B'}||_1 \leq \delta(\epsilon)
\ee
where 
\be
\delta(\epsilon)= 2\sqrt{4\sqrt{2\epsilon}+ \eta(2\sqrt{2\epsilon})} + 2\sqrt{2\epsilon} 
\label{eq:deps}
\ee
and $\eta(x)=-x\log x $. Note, that $\delta(\epsilon)$ vanishes, when $\epsilon$ approaches zero. 
\label{prop:close-to-private} 
}

We then obtain the following corollary. 
{\corollary For $\rho_m$ as defined in $\eqref{eq:rhom}$ there exists a private bit $\gamma_m$, such that $||\gamma_m - \rho_m||\leq \delta(\epsilon)$
with $\delta(\epsilon)= 2\sqrt{4\sqrt{2\epsilon}+ \eta(2\sqrt{2\epsilon})} + 2\sqrt{2\epsilon}$  and 
\be
\epsilon = {2\over 3}(1-(1-{1\over 2^m})^m \times {1\over 1 +{1\over 2^m}}). 
\label{eq:good-eps}
\ee
Note that $\delta= \exp(-O(m))$.
}

\proof
From \cite[(5.18)]{karol-PhD}, we know that by expressing $\rho$ in the form $\rho_m = \sum_{ijkl}|ij\>\<kl|\otimes A_{ijkl}$ we find:
\be
||A_{0011}|| = {1\over 2}(1- {1\over 2^k})^m {1\over 1 + ({1 -2p\over 2p})^m}
\ee
with $k = m$ and $p = {1\over 3}$. Hence $||A_{0011}|| = {1\over 2} - \epsilon$ with $\epsilon ={1\over 2}(1 - (1 - {1\over 2^m})^m\times {1\over {1+{1 \over 2^m}}})$. Thus increasing $\epsilon$ by the multiplicative factor $4\over 3$, we have shown that $\rho_m$ satisfies the assumptions of proposition \ref{prop:close-to-private}.
\qed

We now show how the construction of $\gamma_m$ is done explicitly: Consider the submatrix of the state $\rho_m$ denoted by $A_{0011} = {1\over N_m}[p({\tau_1 - \tau_2\over 2})]^{\otimes m}$ with $N_m = 2(p^m)+ 2({1\over 2} -p)^m$, where $p ={1\over 3}$. Using the singular decomposition, we write $A_{0011} = U^{(00)} \Sigma U^{(11)}$ with $U^{(ii)}$ being unitaries and $\Sigma \geq 0$ a positive operator. Then
\be
\gamma_m = U^{\dagger}_{\tau} [|\psi_-\>\<\psi_-|\otimes (\Tr_{A'B'} U_{\tau}\rho_m U_{\tau}^{\dagger})]U_{\tau}
\label{eq:def-of-gamma-m}
\ee
where $U_{\tau} = \sum_{i} |ii\>\<ii|_{AB}\otimes V_{AB}^{(ii)}$ with $V^{(00)}_{AB} = {U^{(00)}}^{\dagger}$ and $V^{(11)}_{AB} = U^{(11)}$. The idea of the above construction is that by use of a certain {\it twisting} $U_{\tau}$ we can 
decouple $A'B'$ from $AB$ as much as possible and obtain a leftover state on $A'B'$. Replacing the state on the $AB$ system by the singlet state
$|\psi_-\>\<\psi_-|$ and applying the inverse of the twisting $U_{\tau}$ we obtain $\gamma_m$. Note that this state is a private state by construction: it is a "twisted" singlet \cite{pptkey,keyhuge}.

The following lemma provides bounds on the distillable entanglement and the entanglement of formation of the constructed private bit. 
{\lemma For $\gamma_m$ defined in Eq. (\ref{eq:def-of-gamma-m}) satisfies $E_F(\gamma_m) \leq 1 + \exp(-O(m))$ and 
$E_D(\gamma_m) \leq \exp(-O(m))$. 
}

\proof
By construction we have $||\gamma_m - \rho_m||\leq \delta(\epsilon)$, with appropriate $\epsilon$ and $\delta(\epsilon)$. By assumption we have also $E_F(\rho_m) \leq 1$, which, by the  asymptotic continuity of entanglement of formation \cite{BDSW1996}, in formulation of \cite{Synak-asym}, results in
\be
|E_F(\gamma_m) - E_F(\rho_m)| \leq \sqrt{2\delta(\epsilon)}2m^2 \log 2m + \eta(\sqrt{2\delta(\epsilon)})
\ee
provided $\delta(\epsilon)< {1\over 4}$. Since $E_F(\rho_m) \leq 1$ and $\delta(\epsilon)=\exp(-O(m))$ we obtain desired bound.

Now, as it is shown in \cite{Michal2001} we have
\be
E_D(\gamma_m)\leq E_r^{PPT}(\gamma_m),
\label{eq:ed-ptt-bound}
\ee
where $E_r^{PPT}$ is the relative entropy of entanglement distance from the set of states with positive partial transposition. Since this function is 
asymptotically continuous \cite{Synak-asym}, we have 
\be
|E_r^{PPT}(\gamma_m)- E_r^{PPT}(\rho_m)|\leq 4 \eta \log (2m^{2m^2}) + 2 h(\eta)
\ee
with $\eta = ||\gamma_m - \rho_m||$. Since $\rho_m$ is PPT, we have $E_r^{PPT}(\rho_m) = 0$. Thus, we obtain
\be
E_r^{PPT}(\gamma_m) \leq 4\delta(\epsilon)2m^2\log 2m + 2 h(\delta(\epsilon))
\ee
if only $\delta(\epsilon) \leq {1\over 2}$, which, together with (\ref{eq:ed-ptt-bound}) and $\delta(\epsilon)=\exp(-O(m))$, proves the claim.
\qed

Finally, we can prove that $\gamma_m$ has limited key repeater rate. To this end we insert the bounds from the above lemma into theorem \ref{main_theo} and obtain the following statement. 

{\corollary For the private bits $\gamma_m$ defined in Eq. (\ref{eq:def-of-gamma-m}), we have
\be
K_{A\leftarrow C \leftrightarrow B}(\gamma_m)\lesssim {1\over 2}
\ee
in limit of large $m$.
}

\subsection*{On Tightness: A Counterexample for Entanglement Cost}
\label{sec:EoF-counterexample}
Lemmas \ref{th1} and \ref{th2} are new inequalities 
for entanglement measures. It might be worth asking, both from a practical 
and an abstract point of view, whether there are more inequalities of 
that kind for other entanglement measures. 
First, let us note that $E(\tau)\le pE(\r)+(1-p)E(\tilde{\r})$ is trivially 
fulfilled for all LOCC-monotonic measures $E$ and all $0\le p\le1$. 
What would be interesting instead, is a relation of the form
\begin{equation}
  \label{generalBound}
  E(\tau) \le pE_D(\tilde{\r})+(1-p)E(\r)\quad\text{or}\quad E(\tau)\le pE_D(\r)+(1-p)E(\tilde{\r}),
\end{equation}
for some measure $E$ and some weight $p$. If we had a quantum repeater that iterates the 
swapping operation many times, and bound entangled input states, $E$ 
would be reduced by a factor $1-p$ with every step. For measures that 
upper bound the distillable key, such as $E_C$, $E_F$, $E_{sq}$, $E_{sq,c}$, 
$E_R$ or $E_R^\infty$, this would be a significant limitation 
to quantum key repeaters with bound entangled input states. The 
same would hold, if we replaced $E_D$ by $E_N$ or $E_{R,\PPT}$.


We will now show that for $E=E_F$, the entanglement of formation, and $E=E_C$, the entanglement cost, (\ref{generalBound}) cannot hold for all input states. 
Assume that Bob and Charlie apply the following LOCC protocol. Charlie 
performs a generalised Bell state measurement 
$\pro{\Psi^{\n\m}}_{C}$, where 
$\ket{\Psi^{\n\m}}=\frac{1}{\sqrt{d}}\sum_j\omega^{j\n}\ket{j}\te\ket{j+\m}$ 
and $\omega=e^{\frac{2\pi i}{d}}$. 
(Here and in the following the addition is performed modulo $d$.) Charlie then communicates the result $\n,\m$ classically to Alice and Bob. Upon receiving the message, Bob performs 
$U^{\n\m}=\sum_j\omega^{j\n}\ket{j}\bra{j+\m}$. Alice and Bob then store $\m$ classically. 
Charlie's subsystem is then discarded, that is given to Eve.
\begin{proposition}\label{mainProp}
  For the protocol described above, and any $0 < p \leq 1$, 
  there exist states $\r,\tilde{\r}$ such that for $E=E_F$ and $E=E_C$
  \begin{equation}
    E(\tau_{AB}) > p E_D(\tilde{\r}_{C_BB}) + (1-p) E(\r_{AC_A}) \text{ and }E(\tau_{AB}) > p E_D(\r_{AC_A}) + (1-p) E(\tilde{\r}_{C_BB}),
  \end{equation}
  where $\tau$ is the state resulting from the protocol.
\end{proposition}
Our counterexamples are of the form 
$\r_{AB}=\sum_{i,k=0}^{d-1}a_{ik}\ket{ii}\bra{kk}$, which admits a 
purification $\ket{\Phi}_{ABE}=\frac{1}{\sqrt{d}}\sum_i\ket{ii}\te\ket{u_i}$,
where $a_{ik}=\frac{1}{d}\braket{u_k}{u_i}$ and the $\ket{u_i}$ are normalised. 
Such states are called \emph{maximally correlated}. It is easy to see that $\r_A=\r_B=\frac{\1}{d}$. For maximally correlated 
states the entanglement measures involved simplify and $\tau$ can be 
easily calculated. In particular (see \cite{HHHH09} and references therein),
\begin{equation}
  \label{EDmc}
  E_D(\r_{AB}) = E_R(\r_{AB}) = \log{d}-H(\r)
\end{equation}
and
\begin{equation}
  \label{EFmc}
  E_C(\r_{AB})=E_F(\r_{AB}) = \log{d}-I_{\text{acc}}\left(\left\{\frac{1}{d},\ket{u_i}\right\}\right),
\end{equation}
where $I_{\text{acc}}\left(\left\{\frac{1}{d},\ket{u_i}\right\}\right)=\sup_{\{A_j\}\text{POVM}}I(i:j)$ is the \emph{accessible information}. 
Before proceeding with our counterexample for $E_F$ and $E_C$ let us note that (\ref{generalBound}) with $E=E_R$ is trivially fulfilled as for all maximally correlated states $E_D=E_R$.

\begin{lemma}
  \label{maxCorr}
  Let $\r_{AC_A}$ and $\tilde{\r}_{C_BB}$ be maximally correlated, with purifications $\ket{\Phi^1}_{AC_AE_A}=\frac{1}{\sqrt{d}}\sum_i\ket{ii}_{AC_A}\te\ket{u_i}_{E_A}$ 
  and $\ket{\Phi^2}_{C_BBE_B}=\frac{1}{\sqrt{d}}\sum_i\ket{ii}_{C_BB}\te\ket{v_i}_{E_B}$, 
  respectively. Then for every $0<p\le1$, (\ref{generalBound}) with $E=E_F$ or $E=E_C$ implies
  \begin{align}
    \label{IaccBound}
    \frac{1}{d}\sum_\m I_{\text{acc}}\left(\left\{\frac{1}{d},\ket{u_i}\te\ket{v_{i+\m}}\right\}\right)
       &\ge pH(\tilde{\r})\text{ and}   \\
    \frac{1}{d}\sum_\m I_{\text{acc}}\left(\left\{\frac{1}{d},\ket{u_i}\te\ket{v_{i+\m}}\right\}\right)
       &\ge pH(\r).
  \end{align}

\end{lemma}

\proof
Let $0<p\le1$. Let us first show that maximally correlated states preserve their structure under 
the protocol assumed in Proposition \ref{mainProp}. The protocol results in a state $\tau_{AA'BB'}$ purified by
\begin{align}
  \ket{\tilde{\Phi}}
    &=\sum_{\n\m}(\1_{AE_AE_B} \te \pro{\Psi^{\n\m}}_{C} \te U_B^{\n\m})\ket{\Phi^1}_{AC_AE_A}
                                 \te \ket{\Phi^2}_{C_BBE_B} \te\ket{\m\m}_{ab} \ket{\n\m}_{\tilde{E}}\\
    &=\frac{1}{\sqrt{d}}\sum_{\m}\underbrace{\frac{1}{\sqrt{d}}\sum_{i}\ket{ii}_{AB} \te\ket{u_i}_{E_A}\te\ket{v_{i+\m}}_{E_B}}_{=:\ket{\tilde{\Phi}^\m}}\te\ket{\m\m}_{ab}
      \te\underbrace{\frac{1}{\sqrt{d}}\sum_\n \ket{\Psi^{\n\m}}_{C}\te\ket{\n\m}_{\tilde{E}}}_{=:\ket{w_\m}}.
\end{align}
Clearly, $\tau_{AB}^\m:=\tr_{E_AE_B}\pro{\tilde{\Phi}^\m}$ is maximally correlated and $\{\ket{w_\m}\}$ are orthogonal. Therefore Alice and Bobs final state is given by $\tau_{AaBb}=\frac{1}{d}\sum_\m\tau^\m_{AB}\te\pro{\m\m}_{ab}$. By the convexity and LOCC monotonicity of $E_F$, it holds that $E_F(\tau)=\frac{1}{d}\sum_\m E_F(\tau^\m)$. Since we are dealing with maximally correlated states, the same holds true for $E_C$. Now, assume that we have (\ref{generalBound}) with $E=E_F$ or $E=E_C$. Inserting (\ref{EDmc}) and (\ref{EFmc}) into (\ref{generalBound}) gives us 
 \be
    \frac{1}{d}\sum_\m I_{\text{acc}}\left(\left\{\frac{1}{d},\ket{u_i}\te\ket{v_{i+\m}}\right\}\right)
       \ge pH(\tilde{\r})+(1-p)I_{\text{acc}}\left(\left\{\frac{1}{d},\ket{u_i}\right\}\right)
  \ee
and the same for $\r$ and $\ket{v_i}$. Since the accessible information is always non-negative, this implies the Lemma.
\qed

Hence, if we can find an example such that 
$I_{\text{acc}}(\{\frac{1}{d},\ket{u_i}\te\ket{v_{i+\m}}\})<pH(\r)$ and $I_{\text{acc}}(\{\frac{1}{d},\ket{u_i}\te\ket{v_{i+\m}}\})<pH(\tilde{\r})$ 
for all $\m$ we will have Proposition \ref{mainProp}. For this, we make the following ansatz:

\begin{align}
  \ket{\Phi^1}_{AA'C_AC_A'E_A} &= \frac{1}{\sqrt{dn}}\sum_{i=1}^{d}\sum_{j=1}^{n}\ket{ii}_{AC_A}
                                                                 \te\ket{jj}_{A'C_A'}\te U^j\ket{i}_{E_A},\\
  \ket{\Phi^2}_{C_BC_B'BB'E_B} &= \frac{1}{\sqrt{dn}}\sum_{i=1}^{d}\sum_{j=1}^{n}\ket{ii}_{C_BB}
                                                                 \te\ket{jj}_{C_B'B'}\te V^j\ket{i}_{E_B},
\end{align}
where $U^j,V^j$ are unitaries. This is a generalisation of the 
\textit{flower states} introduced in \cite{lock-ent} (see \cite{ChristandlW-lock}). 
Replacing the index $i$ with $(i,j)$, hence also $d$ with $dn$, it is easy to see that those 
are maximally correlated states. Since $\tr_{AA'C_AC_A'}\pro{\Phi^1}=\tr_{C_BC_B'BB'}\pro{\Phi^2}=\frac{\1}{d}$,
we also have $H(\r)=H(\tilde{\r})=\log{d}$. 
Consequently, Proposition \ref{mainProp} follows from Lemma \ref{maxCorr} 
and the next proposition.
\begin{proposition}
  \label{prop1}
 There exists $d_0\in\NN$ such that for all $d\ge d_0$ and $n=d^8$ there are $2n$ unitaries $U^1,\ldots,U^{n},V^1,\ldots,V^{n}$ such that for all $\a=1,\ldots,n$, $\b=1,\ldots,d$,
  \begin{equation}
    I_{\text{acc}}\left(\left\{\frac{1}{dn},U^j\ket{i}_{E_A}\te V^{j+\a}\ket{i+\b}_{E_B}\right\}\right)\le\cO(1).
  \end{equation}
\end{proposition}
Before we can prove Proposition \ref{prop1} we need several technical 
lemmas. Let $n,d\in\NN$.
\begin{lemma}
  \label{Lem7}
  For random unitaries $U^j,V^j,\;j=1,\ldots,n$, $\a\in\{1,\ldots,n\}$, $\b\in\{1,\ldots,d\}$, 
  and $0<\d<\frac{1}{2}$, it holds
  \begin{equation}
    \Pr\left\{\frac{1}{dn}\sum_{i=1}^d\sum_{j=1}^nU^{j}\pro{i}U^{j\dagger}\te V^{j+\a}\pro{i+\b}V^{j+\a\dagger}
                                             \notin\left[\frac{1-\d}{d^2}\1,\frac{1+\d}{d^2}\1\right]\right\}
       \le 2d^2\exp\left(-\frac{n\d^2}{d2\ln2}\right).
  \end{equation}
\end{lemma}
\proof
Let $\a\in\{1,\ldots,n\}$, $\b\in\{1,\ldots,d\}$ and $0<\d<\frac{1}{2}$. Then,
\begin{align}
\EE_{\textbf{UV}}\frac{1}{dn}\sum_{i=1}^d\sum_{j=1}^nU^{j}\pro{i}U^{j\dagger}
          &\te V^{j+\a}\pro{i+\b}V^{j+\a\dagger}                  \\
          &=   \EE_UU\pro{0}U^{\dagger}\te\EE_UU\pro{0}U^{\dagger}
           =\frac{\1}{d^2}, 
\end{align}
so \cite[Thm. 19]{AW02} can be applied, yielding the desired property.
\qed
\begin{lemma}
  \label{lemma2}
  For all $\a\in\{1,\ldots,n\}$, $\b\in\{1,\ldots,d\}$ and 
  $0<\d<\frac{1}{2}$, if $n\ge 6d$ and 
  \begin{equation}
    \frac{1}{dn}\sum_{i=1}^d\sum_{j=1}^nU^{j}\pro{i}U^{j\dagger}\te V^{j+\a}\pro{i+\b}V^{j+\a\dagger}
                                                   \in\left[\frac{1-\d}{d^2}\1,\frac{1+\d}{d^2}\1\right],
  \end{equation}
  then
  \begin{equation}
    I_{\text{acc}}\left(\left\{\frac{1}{dn},U^j\ket{i}_{E_A}\te V^{j+\a}\ket{i+\b}_{E_B}\right\}\right)
                \le\log{dn}-\inf_{\ket{\varphi}}{\tilde{H}^{\a\b}_{\varphi,\d}(\textbf{U},\textbf{V})},
  \end{equation}
  where $\textbf{U}=(U^1,\ldots,U^n)$, $\textbf{V}=(V^1,\ldots,V^n)$ and 
  \begin{equation}
    \tilde{H}^{\a\b}_{\varphi,\d}(\textbf{U},\textbf{V})
      =\sum_{i=1}^d \sum_{j=1}^n \eta\left(\frac{d}{n(1+\d)}
                 \left|\bra{\varphi}_{E_AE_B}U^j\ket{i}_{E_A}\te V^{j+\a}\ket{i+\b}_{E_B}\right|^2 \right),
  \end{equation}
  with $\eta(x)=-x\log x$.
\end{lemma}
\proof
Let $\a\in\{1,\ldots,n\}$, $\b\in\{1,\ldots,d\}$ and $0<\d<\frac{1}{2}$. Without loss of generality, the optimisation in $I_{\text{acc}}$ can be restricted to rank 1 POVMs, hence
\begin{align}
I_{\text{acc}}\left(\left\{\frac{1}{dn},U^j\ket{i}_{E_A}\te V^{j+\a}\ket{i+\b}_{E_B}\right\}\right)
  &=   \sup_{\{\m_k\pro{\varphi_k}\}\text{ rank-1 POVM}}I(ij:k)                                      \\
  &\!\!\!\!\!\!\!\!\!\!\!\!\!\!\!\!\!\!\!\!\!\!\!\!\!\!\!\!\!\!
   =   \log{dn}-\inf_{\{\m_k\pro{\varphi_k}\}}\sum_kp(k)H\bigl(p(ij|k):i=1\ldots d,j=1\ldots n\bigr) \\
  &\!\!\!\!\!\!\!\!\!\!\!\!\!\!\!\!\!\!\!\!\!\!\!\!\!\!\!\!\!\!
   \le \log{dn}-\inf_{\ket{\varphi_k}\in\cH_{E_AE_B}}H\bigl(p(ij|k):i=1\ldots d,j=1\ldots n\bigr),
\end{align}
where
\begin{align}
  p(ijk)  &= \frac{\m_k}{dn}\left|\bra{\varphi_k}U^j\ket{i}\te V^{j+\a}\ket{i+\b}\right|^2, \\
  p(k)    &= \sum_{i=1}^d\sum_{j=1}^np(ijk) \text{ and } p(ij|k) = \frac{p(ijk)}{p(k)}.
\end{align}
By assumption $p(k)\in\left[\frac{(1-\d)\m_k}{d^2},\frac{(1+\d)\m_k}{d^2}\right]$, 
hence
\begin{equation}
  p(ij|k) \ge \frac{d}{n(1+\d)}\left|\bra{\varphi_k}U^j\ket{i}\te V^{j+\a}\ket{i+\b}\right|^2
\end{equation}
and 
\begin{equation}
  p(ij|k) \le \frac{d}{n(1-\d)}\left|\bra{\varphi_k}U^j\ket{i}\te V^{j+\a}\ket{i+\b}\right|^2
          \le \frac{1}{e},
\end{equation}
for $n\ge6d$. As $\eta(x)$ is increasing for $x\le\frac{1}{e}$,
\begin{equation}
  H\bigl(p(ij|k):i=1,\ldots,d,j=1,\ldots,n\bigr)
      \ge \sum_{i=1}^d\sum_{j=1}^n
          \eta\left(\frac{d}{n(1+\d)}\left|\bra{\varphi}U^j\ket{i}\te V^{j+\a}\ket{i+\b}\right|^2\right),
\end{equation}
finishing the proof.
\qed
Next, we lower bound 
$\inf_{\ket{\varphi}}{\tilde{H}^{\a\b}_{\varphi,\d}(\textbf{U},\textbf{V})}$ 
using the following \textit{concentration of measure} result:
\begin{theorem}
  \label{theo6}(Theorem 2.4 in \cite{L05})
  Let $(\cX,g)$ be a compact connected smooth Riemannian manifold with 
  Ricci curvature $\ge  \text{Ric}_{\text{min}}(\cX)>0$ equipped with the normalised Riemannian 
  volume element $d\m=\frac{dv}{V}$. Then for any $\lambda$-Lipschitz 
  function $F$ on $X$ and any $r\ge0$,
  \begin{equation}
    \m\left(\left\{F\le\EE F-r\right\}\right)
        \le \exp{\left(-\frac{\text{Ric}_{\text{min}}(\cX)r^2}{2\lambda^2}\right)}.
  \end{equation}
\end{theorem}
In order to apply Theorem \ref{theo6} we need to lower bound the 
expectation value of $\tilde{H}$.
\begin{lemma}
  \label{lemma3}
There exists $d_1$, such that for $d\ge d_1$, $n=d^8$, $\ket{\varphi}\in\cH_{E_AE_B}$, $\a\in\{1,\ldots,n\}$, $\b\in\{1,\ldots,d\}$ and $\d=\frac{1}{\log{dn}}$
  we have
  \begin{equation}
    \EE_{\textbf{U}\textbf{V}}\tilde{H}^{\a\b}_{\varphi,\d}(\textbf{U},\textbf{V})
                                                      \ge\log{dn}-\cO(1),
  \end{equation}
	where we are using the Haar measure on $\cS\cU(d)^{2n}$.
\end{lemma}
For the proof see Section \ref{appB}.
We also need the fact that $\cS\cU(d)^{2n}$ is a compact connected 
smooth Riemannian manifold with positive Ricci curvature (for 
details see Section \ref{appA}). Next, we need to upper bound 
the Lipschitz constant of $\tilde{H}$ with respect to the Riemannian 
metric of $\cS\cU(d)^{2n}$.
\begin{lemma}
  \label{lemma5}
  For every $n>d\ge8$, $\a\in\{1,\ldots,n\}$, $\b\in\{1,\ldots,d\}$, 
  $0<\d<\frac{1}{2}$ and $\ket{\varphi}\in\cH_{E_AE_B}$, the 
  Lipschitz constant $\tilde{\lambda}$ of $\tilde{H}^{\a\b}_{\varphi,\d}$ 
  is upper bounded
  \begin{equation}
    \tilde{\lambda}\le\frac{8d}{\sqrt{n}}\log{n}.
  \end{equation}
\end{lemma}
The proof can be found in Section \ref{appB}. Apart from applying 
Theorem \ref{theo6} to $\tilde{H}$, we will need the following net result:

\begin{lemma}
  \label{lemma7} (Lemma II.4 in \cite{HLSW04})
  For $0<x<1$ there exists a set $\cM$ of unit vectors in $\cH$ 
  with $|\cM|\le\left(\frac{5}{x}\right)^{2\dim\cH}$ such that for 
  every unit vector $\ket{\varphi}\in\cH$ there exists 
  $\ket{\tilde{\varphi}}\in\cM$ with 
  $\left\|\ket{\varphi}-\ket{\tilde{\varphi}}\right\|_2\le\frac{x}{2}$. 
  Such an $\cM$ is called an ``$x$-net``.
\end{lemma}

Finally, we will need the Lipschitz constant of 
$\hat{H}_{\textbf{UV}}:\cH_{E_AE_B}\rightarrow\RR$, 
$\hat{H}_{\textbf{UV}}(\ket{\varphi})=\tilde{H}^{\a\b}_{\varphi,\d}(\textbf{U},\textbf{V})$.
\begin{lemma}
  \label{lemma8}
  For every $\textbf{U},\textbf{V}$, $n>d\ge8$, $\a\in\{1,\ldots,n\}$, 
  $\b\in\{1,\ldots,d\}$ and $0<\d<\frac{1}{2}$ the Lipschitz constant
  $\hat{\lambda}$ of $\hat{H}_{\textbf{UV}}$ is upper bounded
  \begin{equation}
    \hat{\lambda}\le4\sqrt{2}d\log{n}.
  \end{equation}
\end{lemma}
For the proof see Section \ref{appB}.

\proof \textbf{of Proposition \ref{prop1}}
Let $0< r <1$, $0<\d<\frac{1}{4}$, $d\ge 8$ and 
$n=d^8$. By Lemma \ref{lemma7} there exists an 
$\frac{r}{8\sqrt{2}d\log{n}}$-net $\cM$ of pure states 
in $\cH_{E_AE_B}$ with 
$\left|\cM\right|\le\left(\frac{40\sqrt{2}d\log{n}}{r}\right)^{2d^2}$. 
We will first show that there exists a $d_0$ such that for $d\ge d_0$ there exist $2n$ 
unitaries $U^1,\ldots,U^n,V^1,\ldots,V^n$ fulfilling
\begin{enumerate}[(i)]
  \item $\tilde{H}^{\a\b}_{\tilde{\varphi},\d}(\textbf{UV})
          \ge\EE_{\textbf{UV}}\tilde{H}^{\a\b}_{\tilde{\varphi},\d}-\frac{r}{4}\ 
          \forall\a\in\{1,\ldots,n\},\b\in\{1,\ldots,d\},\ket{\tilde{\varphi}}\in\cM$,
  \item $\frac{1}{dn}\sum_{i=1}^d\sum_{j=1}^nU^{j}\pro{i}U^{j\dagger}\te V^{j+\a}\pro{i+\b}V^{j+\a\dagger}
          \in\left[\frac{1-\d}{d^2}\1,\frac{1+\d}{d^2}\1\right]\;\forall\a\in\{1,\ldots,n\},\b\in\{1,\ldots,d\}$.
\end{enumerate}
Using Theorem \ref{theo6}, 
Lemma \ref{Lem7} and the union bound, we get
\begin{align}
  \Pr\left\{\text{not }(i) \text{ or } \text{not }(ii)\right\} 
     &\le nd\left|\cM\right|\exp{\left(-\frac{cdr^2}{32\tilde{\lambda}^2}\right)}
                                    +2nd^3\exp{\left(-\frac{n\d^2}{2d\ln{2}}\right)}  \\
     &\!\!\!\!\!\!\!\!\!\!\!\!\!\!\!\!\!\!\!\!\!\!\!\!\!\!\!\!\!\!\!\!\!\!\!\!\!\!\!\!
      \le \frac{1}{2}\exp{\left(\left(\ln{4d}+\frac{80\sqrt{2}d^3}{r}\right)8\log{d}
                                       -\frac{cr^2d^{7}}{131072(\log{d})^2}\right)}
           + \frac{1}{2}\exp{\left(\ln{4}+11\ln{d}-\frac{d^{7}\d^2}{2\ln{2}}\right)},
\end{align}
where it has been used that $ \text{Ric}_{\text{min}}(d)=cd$ (see Section \ref{appA}). Both exponents can be made negative for large enough $d_0$ and $d\ge d_0$, implying that 
$\Pr\left\{\text{not }(i) \text{ or } \text{not }(ii)\right\}<1$; 
hence the desired unitaries exist. Now we will show that this implies 
Proposition \ref{prop1}. By (ii) and Lemma \ref{lemma2},
\begin{equation}
  I_{\text{acc}}\left(\left\{\frac{1}{dn},U^j\ket{i}_{E_A}\te V^{j+\a}\ket{i+\b}_{E_B}\right\}\right)
                \le\log{dn}-\inf_{\ket{\varphi}}{\tilde{H}^{\a\b}_{\varphi,\d}(\textbf{U},\textbf{V})}.
\end{equation}
By the definition of the infimum, there exists $\ket{\varphi_0}\in\cH_{E_AE_B}$ 
such that 
$\tilde{H}^{\a\b}_{\varphi_0,\d}(\textbf{U},\textbf{V})
  <\inf_{\ket{\varphi}}{\tilde{H}^{\a\b}_{\varphi,\d}(\textbf{U},\textbf{V})}+\frac{r}{4}$. 
By Lemma \ref{lemma7}, $|\cM|$ contains a state 
$\ket{\tilde{\varphi}_0}$ such that 
$\left\|\ket{\varphi_0}-\ket{\tilde{\varphi}_0}\right\|_2\le\frac{r}{16\sqrt{2}d\log{n}}$. 
By Lemma \ref{lemma8} then, 
\begin{equation}
  \left|\tilde{H}^{\a\b}_{\varphi_0,\d}(\textbf{U},\textbf{V})
        -\tilde{H}^{\a\b}_{\tilde{\varphi}_0,\d}(\textbf{U},\textbf{V})\right|\le\frac{r}{4}.
\end{equation}
Consequently 
$\tilde{H}^{\a\b}_{\tilde{\varphi}_0,\d}(\textbf{U},\textbf{V})
  \le \tilde{H}^{\a\b}_{\varphi_0,\d}(\textbf{U},\textbf{V})+\frac{r}{4}
  <   \inf_{\ket{\varphi}}\tilde{H}^{\a\b}_{\varphi,\d}(\textbf{U},\textbf{V})+\frac{r}{2}$.
Setting $d\ge\max{\{d_0,d_1\}}$ and $\d=\frac{1}{\log{dn}}$, we obtain
\begin{align}
  I_{\text{acc}}\left(\left\{\frac{1}{dn},U^j\ket{i}_{E_A}\te V^{j+\a}\ket{i+\b}_{E_B}\right\}\right)
         &<   \log{dn}-\tilde{H}^{\a\b}_{\tilde{\varphi}_0,\d}(\textbf{U},\textbf{V})+\frac{r}{2}      \\
         &\le \log{dn}-\EE_{\textbf{U},\textbf{V}}\tilde{H}^{\a\b}_{\tilde{\varphi}_0,\d}+\frac{3r}{4} \\
         &\le \cO(1),
\end{align}
where the second and third inequalities are due to (i) and Lemma \ref{lemma3}, respectively.
\qed


\subsection*{Technical Lemmas}
\label{appB}
\label{appA}
We will now briefly review some facts about the Riemannian geometry of the special unitary group. 
\begin{lemma}
  \label{lemmaA} 
  $\cS\cU(d)$, thought of as a sub-manifold in $\CC^{d\times d}$, and equipped 
  with the Hilbert-Schmidt inner product on 
  its tangent spaces, is a compact connected Riemannian manifold.
\end{lemma}
\proof
It is known that $\cS\cU(d)$ is a real semi-simple compact connected 
Lie group \cite{H03}. Every real Lie group is a real smooth manifold. 
Clearly, the Hilbert-Schmidt inner product is a positive definite 
bilinear form. It is also easy to see that it is smooth. Let 
$U\in\cS\cU(d)$ and $X,Y$ be some smooth vector fields on $\cS\cU(d)$, 
that is smooth mappings of $\cS\cU(d)$ into its tangent bundle. 
As it is a composition of smooth maps, the map 
$U\mapsto\tr\left(X(U)^\dagger,Y(U)\right)$ is smooth. Hence the 
Hilbert-Schmidt inner product on the tangent spaces is what is 
referred to as a ``Riemannian metric''. A smooth manifold 
endowed with a Riemannian metric is a Riemannian manifold \cite{D92}. 
\qed
From \cite{B09}, we know that there exists $c>0$ such that
\begin{equation}
    \text{Ric}_{\text{min}}(d):=\inf\text{Ric}(x,x)=cd.
  \end{equation}
  The infimum is taken over all tangent unit vectors and 
  $\text{Ric}$ denotes the Ricci curvature. 


Now we can define a Riemannian distance, which is a metric, 
for $\cS\cU(d)$
\begin{equation}
  d_{\cS\cU(d)}(U,U') = \inf_{\gamma:[0,1]\to\cS\cU(d)
                         \text{ s.t. }\gamma(0)=U,\gamma(1)=U'}\int_0^1\left\|\g'(t)\right\|_{HS}dt.
\end{equation}
The Cartesian product $\cS\cU(d)^{2n}$ is a Riemannian manifold 
as well \cite{L05}. As for its metric, we have
\begin{lemma}\label{lemmaB}
  The Riemannian distance of a Cartesian product $\cM\times\cN$ of 
  Riemannian manifolds is given by the Pythagorean theorem
  \begin{equation}
    d_{\cM\times\cN}((U,V),(\tilde{U},\tilde{V})) = \sqrt{d_{\cM}(U,\tilde{U})^2+d_{\cN}(V,\tilde{V})^2},
  \end{equation}
  for $U,\tilde{U}\in \cM$, $V,\tilde{V}\in \cN$.
\end{lemma}
\proof
We know that for tangent vectors $x,y$, $\|(x,y)\|^2=\|x\|^2+\|y\|^2$. 
We also need the fact that the the length of a curve 
$L(\g)=\int_0^1\left\|\g'(t)\right\|dt$ is independent of the 
parametrisation, that is for an increasing function $\tau:[0,1]\to[0,1]$, 
it holds $L(\g\circ\tau)=L(\g)$. Hence it is always possible to 
find a parametrisation such that $\left\|\g'(t)\right\|$ is constant, 
so $L(\g)=\left\|\g'(t)\right\|$. Consequently,
\begin{align}
  d_{\cM\times\cN}((U,V),(\tilde{U},\tilde{V}))
     &= \inf_{\g\tilde{\g}}\int_0^1\sqrt{\|\g'(t)\|^2+\|\tilde{\g}'(t)\|^2}dt\\
     &= \inf_{\g\tilde{\g}}\sqrt{L(\g)^2+L(\tilde{\g})^2}\\
     &= \sqrt{d_{\cM}(U,\tilde{U})^2+d_{\cN}(V,\tilde{V})^2},
\end{align}
which is what we wanted. 
\qed

The minimum Ricci curvature for a Cartesian product of manifolds is 
just the smallest curvature of the factors. Hence Theorem \ref{theo6} 
can be applied to $\tilde{H}$.

Let us now present the proofs that were omitted in the previous section.

\proof\textbf{of Lemma \ref{lemma3}}
Let $d\ge2$, $n=d^8$, $\ket{\varphi}\in\cH_{E_AE_B},\;\a\in\{1,\ldots,n\}$ and $\b\in\{1,\ldots,d\}$. 
We need to lower bound $\EE\tilde{H}$. For a probability distribution 
$\{p_i\}$ it holds that 
$H_2(p)=-\log{\left(\sum_ip_i^2\right)}\le\sum_i\eta(p_i)=H(p)$. 
Here, however, we have 
$\tilde{p}_{ij}=\frac{d}{n(1+\d)}\left|\bra{\varphi}U^j\ket{i}\te V^{j+\a}\ket{1+\b}\right|^2$. Note that $0\le\tilde{p}_{ij}\le\frac{d}{n}\le\frac{1}{e}$. The $\{\tilde{p}_{ij}\}$ are, in general, no probability distribution. However, Lemma \ref{Lem7} tells us that they are most likely close to one. Namely, for $0<\d<\frac{1}{4}$,
\begin{equation}
P\left(\sum_{i=1}^d\sum_{j=1}^n\tilde{p}_{ij}\notin\left[\frac{1-\d}{1+\d},1\right]\right)\le2d^2\exp\left(-\frac{n\d^2}{d\;2\ln{2}}\right).
\end{equation} 
In order to stop $H_2$ from diverging, let us add a little perturbation that keeps $\tilde{p}_{ij}$ away from $0$. Namely, we define
\begin{equation} 
\hat{p}_{ij}=(1-\e)\tilde{p}_{ij}+\e\frac{1}{dn}.
\end{equation}
By concavity and monotonicity of $\eta$ on $[0,\frac{1}{e}]$,
\begin{equation}
\eta(\hat{p}_{ij})\le\eta((1-\e)\tilde{p}_{ij})+\eta\left(\frac{\e}{nd}\right)\le\eta(\tilde{p}_{ij})+\eta\left(\frac{\e}{nd}\right).
\end{equation}
Hence, choosing $\e=\frac{1}{\log{dn}}$, we obtain $H(\tilde{p})\ge H(\hat{p})-\cO(1)$. Next, let us note that if $\sum_{ij}\tilde{p}_{ij}\in\left[\frac{1-\d}{1+\d},1\right]$, it also holds $\sum_{ij}\hat{p}_{ij}\in\left[\frac{1-\d}{1+\d},1\right]$. Let us call this event $G$. If $G$ is true, by Jensen's inequality,
\begin{equation}
H(\hat{p})\ge\sum_{ij}\hat{p}_{ij}H_2(\hat{p})-\eta\left(\sum_{ij}\hat{p}_{ij}\right)\ge\frac{1-\d}{1+\d}H_2(\hat{p})-\eta\left(\frac{1-\d}{1+\d}\right).
\end{equation}
Hence,
\begin{align}
\EE_{\textbf{UV}}H(\tilde{p})&\ge\EE_{\textbf{UV}}H(\hat{p})-\cO(1)\\
&\ge\int_{G}d\textbf{UV}\;H(\hat{p})-\cO(1)\\
&\ge\frac{1-\d}{1+\d}\int_{G}d\textbf{UV}\;H_2(\hat{p})-\cO(1)\\
&=\frac{1-\d}{1+\d}\left(\EE_{\textbf{UV}}H_2(\hat{p})-\int_{\textbf{UV}\notin G}d\textbf{UV}\;H_2(\hat{p})\right)-\cO(1)\\
&\ge\frac{1-\d}{1+\d}\left(\EE_{\textbf{UV}}H_2(\hat{p})-2d^2\exp\left(-\frac{n\d^2}{d\;2\ln{2}}\right)\log{\frac{dn}{\e^2}}\right)-\cO(1),
\end{align}
so it is sufficient to lower bound the expectation value of $H_2(\hat{p})$.
\begin{align}
 \EE_{\textbf{U}\textbf{V}}H_2(\hat{p})&\ge-\log{\left(\EE_{\textbf{U}\textbf{V}}\sum_{ij}\hat{p}_{ij}^2\right)}\\
&=-\log{\left(nd\;\EE_{UV}\hat{p}^2_{00}\right)}\\
&=-\log{\left(nd\left((1-\e)^2\EE_{UV}\tilde{p}^2_{00}+\frac{2\e(1-\e)}{nd}\EE_{UV}\tilde{p}_{00}+\frac{\e^2}{n^2d^2}\right)\right)},
\end{align}
where
\begin{align}
\EE_{UV}\tilde{p}_{00}&\le\frac{d}{n}\EE_{UV}\tr\left(\pro{\varphi}U\pro{0}U^\dagger\te V\pro{0}V^\dagger\right)\\
&=\frac{d}{n}\tr\left(\pro{\varphi}(\EE_UU\pro{0}U^\dagger)^{\te2}\right)\\
&=\frac{1}{nd}
\end{align}
and, using a 2-design,
\begin{align}
\EE_{UV}\tilde{p}^2_{00}&\le\frac{d^2}{n^2}\EE_{UV}\tr\left(\pro{\varphi}U\pro{0}U^\dagger\te V\pro{0}V^\dagger\right)^2\\
&=\frac{d^2}{n^2}\tr\left(\pro{\varphi}^{\te2}\left((\EE_UU\pro{0}U^\dagger)^{\te2}\right)^{\te2}\right)\\
&=\frac{4}{n^2(d+1)^2}\tr\left(\pro{\varphi}_{E_AE_B}^{\te2}\P^+_{E_AE_A}\te\P^+_{E_BE_B}\right)\\
&\le\frac{4}{n^2d^2},
\end{align}
where $\P^+$ denotes the projector onto the symmetric subspace. Hence,
\begin{equation}
 \EE_{\textbf{U}\textbf{V}}H_2(\hat{p})\ge\log{nd}-\log{\left(4(1-\e)^2+2(1-\e)\e+\e^2\right)}\ge\log{nd}-\log{7}.
\end{equation}
Choosing $\d=\frac{1}{\log{dn}}$, for large enough $d_1$ and $d\ge d_1$ we obtain
\begin{equation}
\EE_{\textbf{U}\textbf{V}}\tilde{H}^{\a\b}_{\varphi,\d}(\textbf{U},\textbf{V})=\EE_{\textbf{U}\textbf{V}}H(\tilde{p})\ge\log{dn}-\cO(1),
\end{equation}
and we are done.
\qed

\medskip
Before proving Lemma \ref{lemma5}, we need to upper bound the Lipschitz constant of the function $H_{\b\d}':\bigoplus_{j=1}^n\cH_{E_AE_B}\to\RR$,
\begin{equation}
H_{\b\d}'(\ket{\phi_1},\ldots,\ket{\phi_n})=\sum_{i=1}^d\sum_{j=1}^n\eta\left(\frac{d}{n(1+\d)}\tr\left(\pro{i}\te\pro{i+\b}\pro{\phi_j}\right)\right).
\end{equation} 
Note that for $\ket{\phi_j}=U^{j\dagger}\te V^{j+\a\dagger}\ket{\varphi}$, 
\begin{equation}
\tilde{H}^{\a\b}_{\varphi\d}(\textbf{U}\textbf{V})=H_{\b\d}'(U^{1\dagger}\te V^{1+\a\dagger}\ket{\varphi},\ldots,U^{n\dagger}\te V^{n+\a\dagger}\ket{\varphi}).
\end{equation}
\begin{lemma}\label{lemma4}
For all $n>d\ge8$, $0<\d<\frac{1}{2}$, $\b\in\{1,\ldots,d\}$ the Lipschitz constant $\lambda'$ of $H_{\b\d}'$ is upper bounded 
\begin{equation}
\lambda'\le\frac{4\sqrt{2}d}{\sqrt{n}}\log{n}.
\end{equation}
\end{lemma}
\proof
Let $n>d\ge8$, $0<\d<\frac{1}{2}$ and $\b\in\{1,\ldots,d\}$. We will make use of the fact that $\lambda'^2=\sup_{\braket{\phi_j}{\phi_j}\le1\forall j}\nabla H_{\b\d}'\cdot\nabla H_{\b\d}'$. Writing $\ket{\phi_j}=\sum_{lm=1}^d\phi_{l,m}^{(j)}\ket{lm}$, we get 
\begin{equation}
H_{\b\d}'(\ket{\phi_1},\ldots,\ket{\phi_n})=\sum_{i=1}^d\sum_{j=1}^n\eta\left(\frac{d}{n(1+\d)}\left|\phi_{i,i+\b}^{(j)}\right|^2\right)=\sum_{i=1}^d\sum_{j=1}^n\eta\left(cr_{ij}^2\right),
\end{equation}
where we have defined $b=\frac{d}{n(1+\d)}$ and $r_{ij}=\left|\phi_{i,i+\b}^{(j)}\right|$. By assumption $b<1$. Computing the gradient we obtain
\begin{align}
\sup_{\braket{\phi_j}{\phi_j}\le1\forall j}\nabla H_{\b\d}'\cdot\nabla H_{\b\d}'&=\sup_{\braket{\phi_j}{\phi_j}\le1\forall j}\frac{4b}{(\ln{2})^2}\sum_{i=1}^d\sum_{j=1}^nbr^2_{ij}\left(\ln{(br_{ij}^2)}+1\right)^2\\
&\le\sup_{\sum_{i=1}^dr_{ij}^2\le1\forall j}\frac{4b}{(\ln{2})^2}\left(\sum_{i=1}^d\sum_{j=1}^nbr^2_{ij}\left(\ln{br^2_{ij}}\right)^2+bn\right)\\
&=\frac{4bn}{(\ln{2})^2}\left(\sup_{\sum_{i=1}^dy_i\le b,\;y_i\ge0\forall i}\sum_{i=1}^dy_i(\ln{y_i})^2+b\right)
\end{align}
Using Lagrange multipliers, it can be shown that for $d\ge8$ the maximum is attained at $y_i=\frac{b}{d}$, hence
\begin{equation}
\lambda'^2\le\frac{4b^2n}{(\ln{2})^2}\left(\left(\ln{\frac{b}{d}}\right)^2+1\right)\le\frac{32d^2}{n}\left(\log{n}\right)^2,
\end{equation}
finishing the proof.
\qed

\medskip
\proof\textbf{of Lemma \ref{lemma5}}
Let $U_1,\ldots,U_n,V_1,\ldots,V_n,U'_1,\ldots,U'_n,V'_1,\ldots,V'_n\in\cS\cU(d)$. Then
\begin{align}
\left|\tilde{H}^{\a\b}_{\varphi\d}(\textbf{U},\textbf{V})-\tilde{H}^{\a\b}_{\varphi\d}(\textbf{U}',\textbf{V}')\right|&\le\lambda'\left\|\bigoplus_{j=1}^n\left(U^\dagger_j\te V^\dagger_{j+\a}-U'^\dagger_j\te V'^\dagger_{j+\a}\right)\ket{\varphi}\right\|_2\\
&=\lambda'\sqrt{\sum_{j=1}^n\left\|\left(U^\dagger_j\te V^\dagger_{j+\a}-U'^\dagger_j\te V'^\dagger_{j+\a}\right)\ket{\varphi}\right\|_2^2}\\
&\le\lambda'\sqrt{\sum_{j=1}^n\left\|\left(U^\dagger_j\te V^\dagger_{j+\a}-U'^\dagger_j\te V'^\dagger_{j+\a}\right)\right\|_\infty^2}\\
&\le \sqrt{2}\lambda'\sqrt{\sum_{j=1}^n\left\|U_j-U'_j\right\|_\infty^2+\sum_{j=1}^n\left\|V_j-V'_j\right\|_\infty^2}.
\end{align}
Since
\begin{align}
d_{\text{Riem}}(U,U')&=\inf_\gamma\int_a^b\left\|\g'(t)\right\|_{HS}dt\ge\inf_\gamma\left\|\int_a^b\g'(t)dt\right\|_{HS}\\
&=\inf_\gamma\left\|\g(a)-\g(b)\right\|_{HS}=\left\|U-U'\right\|_{HS}\ge\left\|U-U'\right\|_{\infty},
\end{align}
we get $\tilde{\lambda}=\sqrt{2}\lambda'$. Applying Lemma \ref{lemma4} finishes the proof.
\qed

\medskip
\proof\textbf{of Lemma \ref{lemma8}}
Let $\textbf{U},\textbf{V}\in\cS\cU(d)^d,\;\a\in\{1,\ldots,n\},\;\b\in\{1,\ldots,d\}$. 
Then for all $\ket{\varphi},\ket{\varphi'}\in\cH$,
\begin{align}
  \left|\hat{H}_{\textbf{UV}}(\ket{\varphi})-\hat{H}_{\textbf{UV}}(\ket{\varphi'})\right|
    &\le\lambda'\left\|\bigoplus_{j=1}^nU^j\te V^{j+\a}\left(\ket{\varphi}-\ket{\varphi'}\right)\right\|_2\\
    &=  \lambda'\sqrt{\sum_{j=1}^n\left\|U^j\te V^{j+\a}\left(\ket{\varphi}-\ket{\varphi'}\right)\right\|_2^2}\\
    &=\lambda'\sqrt{n}\left\|\ket{\varphi}-\ket{\varphi'}\right\|_2,
\end{align}
where we have used that the Hilbert space norm is unitarily invariant.
\qed

\pagebreak\section*{Supplementary Note 6} \label{sub:erasure}


\subsection*{Replacing distillable entanglement by (one-way) non-distillable entanglement}

In contrast to the limitations on quantum key repeaters described in the earlier sections, this section shows that in some cases the use of a large amount of distillable entanglement in the form of EPR states can be replaced by one-way non-distillable states. 

In order to see this, consider a situation in which Alice and Charlie share a private bit $\gamma_{AA'C_AC_A'}$, which is almost PPT in the sense that $E_N(\gamma)\leq \epsilon$. This implies that the shield dimension $|C_A'|=d \gtrapprox \frac{1}{\epsilon}$: we write $\gamma$ in its $X$-form and calculate
\be
E_N(||\gamma^\Gamma||)=\log (||\sqrt{X^\dagger X}^\Gamma||_1+||X^\Gamma||_1)\geq \log (1+||X^\Gamma||_1)\gtrapprox ||X^\Gamma||_1
\ee
which holds for small log negativity. $d \gtrapprox \frac{1}{\epsilon}$ now follows, since $||X^\Gamma||\geq \frac{1}{d}$ for $||X||_1=1$ (the diamond norm of the transpose map in dimension $d$ equals $d$). Applying the standard quantum repeater protocol based on teleportation would thus require Charlie and Bob to share $1+ \log d$ EPR pairs. 

Instead let now Charlie and Bob share only one EPR pair $\proj{\phi}_{C_BB}$ and a copy of the Choi-Jamilkowski state corresponding to the 50\% erasure channel: $\rho_{C_B'B'}=\frac{1}{2}\proj{\psi}+ \frac{1}{2} \frac{\mathbf{1}}{d}\otimes \proj{e}$, where $\ket{\psi}=\frac{1}{\sqrt{d}}\sum_i^d \ket{ii}$ and $\ket{e}$ is the erasure symbol orthogonal to $\{\ket{i} \}$. We emphasize that the one-way (from Charlie to Bob) distillable key rate and hence also the corresponding rate of distillable entanglement vanish for this state as it admits a symmetric extension.

Now let Charlie teleport system $C_A$ to Bob by use of the EPR pair and $C_A'$ by using $\rho$ instead of $\proj{\psi}$. It is easy to verify that the resulting state has the form
\be
\sigma_{AA'BB'}= \frac{1}{2}\gamma_{AA'BB'}+ \frac{1}{2}\gamma_{AA'B}\otimes \proj{e},
\ee
where $\gamma_{AA'B}=\tr_{B'}\gamma_{AA'BB'}$.
In order to compute a lower bound on the key rate of this state, we will convert it into a cqq state: Consider a purification $\sigma_{AA'BB'E}$. Let Alice measure her key system in the computational basis with outcome stored in register $X$ and let both players remove (but keep in their labs) the shield systems. The resulting state has the form
\be
\sigma_{XBE}= \frac{1}{2}(\proj{00}+ \proj{11})\otimes \gamma_E + \frac{1}{2}(\proj{00}\otimes \sigma_{0,E}+ \proj{11}\otimes  \sigma_{1,E})
\ee
for certain states $\gamma_E, \sigma_{0, E}, \sigma_{1, E}$ of Eve. It is now easy to compute the lower bound on the one-way (from Alice to Bob) key rate $K^\rightarrow(\sigma_{XBE})$ given by Devetak and Winter \cite{DevetakWinter-hash}: $I(X:B)_\sigma-I(X:E)_\sigma \geq \frac{1}{2}$. In conclusion, a constant key rate can be obtained with a single EPR pair and the (one-way) non-distillable erasure channel. 

\pagebreak\section*{Supplementary References}



\end{document}